\pdfoutput=1
\documentclass{aa}  
\usepackage{xcolor}
\usepackage{graphicx}
\usepackage{siunitx}
\usepackage[breaklinks]{hyperref}
\hypersetup{colorlinks,citecolor=blue}
\usepackage{array}
\usepackage{upgreek}

\usepackage{url}
\usepackage{txfonts}
\graphicspath{{./figures/}}

\begin{document} 

   \title{Protostellar accretion traced with chemistry}
   \subtitle{High resolution C$^{18}$O and continuum observations towards deeply embedded protostars in Perseus}

   \author{S{\o}ren Frimann\inst{\ref{inst1}}\thanks{Current address: Institut de Ci\`{e}ncies del Cosmos, Universitat de Barcelona, IEEC-UB, Mart\'{i} Franqu\`{e}s 1, E-08028 Barcelona, Spain}
          \and
          Jes K. J\o rgensen\inst{\ref{inst1}}
          \and
          Michael M. Dunham\inst{\ref{inst2}}
          \and
          Tyler L. Bourke\inst{\ref{inst3}}
          \and
          Lars E. Kristensen\inst{\ref{inst1}}
          \and
          Stella S. R. Offner\inst{\ref{inst4}}
          \and
          Ian W. Stephens\inst{\ref{inst5}}
          \and
          John J. Tobin\inst{\ref{inst6},\ref{inst7}}
          \and
          Eduard I. Vorobyov\inst{\ref{inst8},\ref{inst9}}
          }

   \institute{Centre for Star and Planet Formation, Niels Bohr Institute and Natural History Museum of Denmark, University of Copenhagen, \O ster Voldgade 5-7, DK-1350 Copenhagen K, Denmark \\ \email{sfrimann@icc.ub.edu} \label{inst1}
   \and
   Department of Physics, SUNY Fredonia, Fredonia, New York 14063, USA \label{inst2}
   \and
   SKA Organization, Jodrell Bank Observatory, Lower Withington, Macclesfield, Cheshire SK11 9DL, UK \label{inst3}
   \and
   Department of Astronomy, University of Massachusetts, Amherst, MA 01003 USA \label{inst4}
   \and
   Harvard-Smithsonian Center for Astrophysics, Cambridge, MA 02138, USA \label{inst5}
   \and
   Homer L.\ Dodge Department of Physics and Astronomy, University of Oklahoma, 440 W.\ Brooks Street, Norman, OK 73019, USA \label{inst6}
   \and
   Leiden Observatory, Leiden University, P.O.\ Box 9513, 2300-RA Leiden, The Netherlands \label{inst7}
   \and
   Department of Astrophysics, The University of Vienna, Vienna, 1180, Austria \label{inst8}
   \and
   Research Institute of Physics, Southern Federal University, Rostov-on-Don 344090, Russia \label{inst9}
}
   \date{Received; accepted}
 
  \abstract
   {Understanding how accretion proceeds is a key question of star formation, with important implications for both the physical and chemical evolution of young stellar objects. In particular, very little is known about the accretion variability in the earliest stages of star formation.}
   {To characterise protostellar accretion histories towards individual sources by utilising sublimation and freeze-out chemistry of CO.}
   {A sample of 24 embedded protostars are observed with the Submillimeter Array (SMA) in context of the large program ``Mass Assembly of Stellar Systems and their Evolution with the SMA'' (MASSES). The size of the C$^{18}$O emitting region, where CO has sublimated into the gas-phase, is measured towards each source and compared to the expected size of the region given the current luminosity. The SMA observations also include \SI{1.3}{mm} continuum data, which are used to investigate whether a link can be established between accretion bursts and massive circumstellar disks.}
   {Depending on the adopted sublimation temperature of the CO ice, between \SI{20}{\percent} and \SI{50}{\percent} of the sources in the sample show extended C$^{18}$O emission indicating that the gas was warm enough in the past that CO sublimated and is currently in the process of refreezing; something which we attribute to a recent accretion burst. Given the fraction of sources with extended C$^{18}$O emission, we estimate an average interval between bursts of \SIrange[range-phrase=--]{20000}{50000}{yr}, which is consistent with previous estimates. No clear link can be established between the presence of circumstellar disks and accretion bursts, however the three closest known binaries in the sample (projected separations <\SI{20}{AU}) all show evidence of a past accretion burst, indicating that close binary interactions may also play a role in inducing accretion variability.}
  {}

   \maketitle
%

\newcommand{\ramses}{\texttt{RAMSES}\xspace}
\newcommand{\radmc}{\texttt{RADMC-3D}\xspace}
\newcommand{\co}{C$^{18}$O\xspace}

\newcommand{\jor}{JVWB15\xspace}

\section{Introduction}
\label{sec:introduction}

One of the key questions of star formation regards the way in which young stellar objects (YSOs) obtain their mass. Specifically, are the accretion rates onto YSOs best characterised by a smooth decline from early to late stages or by intermittent bursts of high accretion? The question of how accretion proceeds is a central one because it holds important clues about the physical and chemical evolution of YSOs such as condensation of chondrules \citep{Boley:2008gk} and the formation of complex organic molecules \citep{Taquet:2016ge}. While there is strong evidence of episodic accretion in the more evolved pre-main-sequence stars (see \citealt{Audard:2014gb} for a recent review), it remains challenging to determine how accretion proceeds in the earliest stages of star formation. In this paper, we model the sublimation and freeze-out of CO to study the accretion histories in a sample of 24 deeply embedded protostars in the Perseus molecular cloud.

\begin{table*}
\caption{Sample of sources.}
\label{tbl:object}
\centering
\begin{tabular}{l l c c S S[separate-uncertainty,table-format=1.2(2)] S[omit-uncertainty] c}
\hline\hline
\multicolumn{2}{c}{Source ID\tablefootmark{a}} & {$\alpha_\mathrm{J2000}$\tablefootmark{b}} & {$\delta_\mathrm{J2000}$\tablefootmark{b}} & {$v_0$\tablefootmark{c}} & {$L_\mathrm{bol}$} & {$T_\mathrm{bol}$} & Reference\tablefootmark{d} \\
           &            & (hh mm ss.s) & (dd mm ss.s) & {(\si{km.s^{-1}})} & {($L_\sun$)} & {(K)} & \\
\hline
Per-emb  1 & HH 211     & 03 43 56.5 & 32 00 52.9 &  9.2 & 1.8(1)   & 27(1)    & \citet{Sadavoy:2014jt} \\
Per-emb  2 & IRAS 03292 & 03 32 18.0 & 30 49 47.6 &  5.1 & 0.90(7)  & 27(1)    & \citet{Sadavoy:2014jt} \\
Per-emb  5 & IRAS 03282 & 03 31 21.0 & 30 45 30.2 &  5.5 & 1.3(1)   & 32(2)    & \citet{Sadavoy:2014jt} \\
Per-emb 11 & IC348-MMS  & 03 43 56.9 & 32 03 04.6 &  9.0 & 1.5(1)   & 30(2)    & \citet{Sadavoy:2014jt} \\
Per-emb 12 & IRAS 4A    & 03 29 10.5 & 31 13 31.0 &  6.6 & 7.0(7)   & 29(2)    & \citet{Sadavoy:2014jt} \\
Per-emb 13 & IRAS 4B    & 03 29 12.0 & 31 13 01.5 &  5.3 & 4.0(3)   & 28(1)    & \citet{Sadavoy:2014jt} \\
Per-emb 14 & IRAS 4C    & 03 29 13.5 & 31 13 58.0 &  7.6 & 0.70(8)  & 31(2)    & \citet{Sadavoy:2014jt} \\
Per-emb 16 &            & 03 43 51.0 & 32 03 24.8 &  8.5 & 0.40(4)  & 39(2)    & \citet{Sadavoy:2014jt} \\
Per-emb 18 &            & 03 29 11.3 & 31 18 31.3 &  8.0 & 3.6(5)   & 46(3)    & \citet{Sadavoy:2014jt} \\
Per-emb 19 &            & 03 29 23.5 & 31 33 29.5 &  7.5 & 0.36(5)  & 60(3)    & \citet{Enoch:2009ch}   \\
Per-emb 21 &            & 03 29 10.7 & 31 18 20.5 &  8.8 & 3.6(5)   & 46(3)    & \citet{Sadavoy:2014jt} \\
Per-emb 22 & L1448 IRS2 & 03 25 22.3 & 30 45 14.0 &  3.7 & 3.6(5)   & 43(2)    & \citet{Sadavoy:2014jt} \\
Per-emb 25 &            & 03 26 37.5 & 30 15 28.0 &  5.6 & 0.95(2)  & 68(12)   & \citet{Enoch:2009ch}   \\
Per-emb 26 & L 1448 C   & 03 25 38.8 & 30 44 06.3 &  5.1 & 9.2(13)  & 47(2)    & \citet{Sadavoy:2014jt} \\
Per-emb 27 & IRAS 2A    & 03 28 55.6 & 31 14 36.6 &  6.3 & 19.0(4)  & 69(1)    & \citet{Enoch:2009ch}   \\
Per-emb 28 &            & 03 43 51.0 & 32 03 07.9 &  8.5 & 0.70(8)  & 45(2)    & \citet{Sadavoy:2014jt} \\
Per-emb 29 & B1-c       & 03 33 17.9 & 31 09 32.0 &  3.0 & 3.7(4)   & 48(1)    & \citet{Sadavoy:2014jt} \\
Per-emb 33 & L 1448 N   & 03 25 36.5 & 30 45 22.3 &  4.9 & 8.3(8)   & 57(3)    & \citet{Sadavoy:2014jt} \\
Per-emb 35 & IRAS 1     & 03 28 37.1 & 31 13 30.7 &  5.5 & 9.1(3)   & 103(26)  & \citet{Enoch:2009ch}   \\
Per-emb 42 &            & 03 25 39.1 & 30 43 58.0 &  5.5 & 0.68(85) & 163(51)  & \citet{Enoch:2009ch}   \\
Per-emb 44 & SVS 13A    & 03 29 03.8 & 31 16 03.7 &  8.6 & 32.5(71) & 188(9)   & \citet{Enoch:2009ch}   \\
Per-emb 47 & IRAS 03254 & 03 28 34.5 & 31 00 51.1 &  7.6 & 1.2(1)   & 230(17)  & \citet{Enoch:2009ch}   \\
Per-emb 53 & B5-IRS1    & 03 47 41.6 & 32 51 43.9 & 10.3 & 4.7(9)   & 287(8)   & \citet{Enoch:2009ch}   \\
Per-emb 61 &            & 03 44 21.3 & 31 59 32.6 &  9.6 & 0.24(16) & 371(107) & \citet{Enoch:2009ch}   \\
\hline
\end{tabular}
\tablefoot{
\tablefoottext{a}{Per-emb~XX names follow the naming scheme of \citet{Enoch:2009ch}.}
\tablefoottext{b}{Positions from the Spitzer cores to disks survey \citep{Evans:2009bk}.}
\tablefoottext{c}{Systemic velocities based on Gaussian fits to C$^{18}$O spectra (see Sect.~\ref{sec:observations}).}
\tablefoottext{d}{Luminosity and $T_\mathrm{bol}$ reference.}
}
\end{table*}

Evidence of episodic accretion in YSOs include the FU~Orionis objects (FUors), which are pre-main-sequence stars showing optical luminosity bursts (\citealt{Herbig:1966jo,Herbig:1977gf}). Theoretically, such outbursts can be tied to accretion instabilities in the circumstellar disk encircling the YSO (e.g.\ \citealt{Bell:1994cp,Armitage:2001jl,Vorobyov:2005kv,Zhu:2009fv}), leading to a lot of material being dumped onto the central object over a short period of time. It is not known if deeply embedded protostars undergo episodic accretion events to the same degree as their older more evolved counterparts, since searches for accretion bursts are most easily carried out at optical and near-infrared wavelengths where the deeply embedded objects are not detected. So far, the only direct detection of a burst in a deeply embedded object is HOPS~383, a Class 0 protostar in Orion, which increased its \SI{24}{\micro\metre} flux by a factor of 35 between 2004 and 2008 \citep{Safron:2015ew}. 

Because of the challenges associated with the direct detection of episodic accretion events in deeply embedded protostars, indirect methods have been used to gather evidence of variable accretion during the embedded phase. Examples include the Class~0 protostar, L673-7, whose integrated outflow properties were used to show that its accretion rate must have been significantly higher in the past \citep{Dunham:2010fk}. Furthermore, observations of knots in the outflows of some protostars, which can be attributed to variations in the accretion rate, can be used to estimate the interval between episodic accretion events (e.g.\ \citealt{Bachiller:1990va,Bachiller:1991ui,Lee:2009gp,Arce:2013kt,Plunkett:2015in}), with studies reporting burst intervals ranging from approximately \SIrange{100}{6000}{yr}.

It is also possible to use chemistry to probe the accretion histories of protostars. During an accretion burst, molecules that are normally frozen-out onto dust grains sublimate into the gas-phase because of the enhanced protostellar heating. Once the burst ends, the envelope cools rapidly \citep{Johnstone:2013je}, whereas the time scale for the molecules to refreeze back onto the dust grains is in the range \SIrange{e3}{e5}{yr} for typical envelope densities ranging from \SIrange{e7}{e5}{cm^{-3}} \citep{Rodgers:2003dg}. Observations of this out-of-equilibrium state, which manifests itself by stronger line intensities extending over larger areas of spatially resolved emission, may be used as a method to detect past accretion bursts \citep{Lee:2007bq,Visser:2012dp,Vorobyov:2013kq,Visser:2015ew}. It is also possible to constrain the accretion history of the protostar by looking at absorption bands of interstellar ices. For example, pure CO$_2$ ice is believed to form on dust grains during the sublimation of CO ice \citep{Lee:2007bq}, which indicates temperatures $\gtrsim$\,\SIrange[range-phrase=--]{20}{30}{K}. Observations of pure CO$_2$ ice absorption bands towards low-luminosity protostars have been used to show that some sources have undergone significant thermal processing in the past (e.g.\ \citealt{Kim:2012hz,Poteet:2013cq}).

\begin{table*}
\caption{Observing log.}
\label{tbl:obs2}
\centering
\begin{tabular}{@{\extracolsep{2pt}}l c c S[table-format=3] S[table-format=2.2] c c@{}}
\hline\hline
Source & SUB date\tablefootmark{a} & EXT date\tablefootmark{a} & \multicolumn{2}{c}{RMS\tablefootmark{b}} & \multicolumn{2}{c}{Synth.\ beam (P.A.)\tablefootmark{c}} \\
\cline{4-5}\cline{6-7}
 & & & {C$^{18}$O} & {Cont.} & C$^{18}$O & Cont. \\
\hline
Per-emb 1   & 07 Dec 2014 & ...         & 140 &  1.78 & \ang{;;4.46} $\times$ \ang{;;3.34} (\ang{-13.6}) & \ang{;;4.37} $\times$ \ang{;;3.26} (\ang{-13.7}) \\
Per-emb 2   & 22 Nov 2014 & 06 Sep 2014 & 125 &  2.08 & \ang{;;3.09} $\times$ \ang{;;2.81} (\ang{-1.8})  & \ang{;;4.30} $\times$ \ang{;;3.35} (\ang{-17.1}) \\
Per-emb 5   & 22 Nov 2014 & 06 Sep 2014 &  88 &  1.33 & \ang{;;3.12} $\times$ \ang{;;2.73} (\ang{-10.8}) & \ang{;;4.28} $\times$ \ang{;;3.36} (\ang{-17.4}) \\
Per-emb 11  & 07 Dec 2014 & ...         &  84 &  1.93 & \ang{;;4.46} $\times$ \ang{;;3.33} (\ang{-14.9}) & \ang{;;4.37} $\times$ \ang{;;3.27} (\ang{-14.8}) \\
Per-emb 12  & 13 Dec 2014 & ...         & 174 & 15.50 & \ang{;;4.22} $\times$ \ang{;;3.40} (\ang{-9.5})  & \ang{;;4.06} $\times$ \ang{;;3.56} (\ang{9.4})   \\
Per-emb 13  & 20 Nov 2014 & 04 Sep 2014 &  95 &  5.95 & \ang{;;2.61} $\times$ \ang{;;2.36} (\ang{20.3})  & \ang{;;4.15} $\times$ \ang{;;3.25} (\ang{-12.4}) \\
Per-emb 14  & 13 Dec 2014 & ...         & 152 &  4.80 & \ang{;;4.12} $\times$ \ang{;;3.34} (\ang{-9.6})  & \ang{;;4.06} $\times$ \ang{;;3.56} (\ang{7.0})   \\
Per-emb 16/28 & 07 Dec 2014 & ...         & 117 &  1.61 & \ang{;;4.15} $\times$ \ang{;;3.45} (\ang{-5.9})  & \ang{;;4.35} $\times$ \ang{;;3.26} (\ang{-14.2}) \\
Per-emb 18/21 & 27 Nov 2014 & ...         & 189 &  2.66 & \ang{;;4.59} $\times$ \ang{;;2.95} (\ang{-25.4}) & \ang{;;4.86} $\times$ \ang{;;3.31} (\ang{-27.3}) \\
Per-emb 19  & 14 Dec 2014 & ...         &  96 &  1.23 & \ang{;;4.02} $\times$ \ang{;;3.33} (\ang{-6.0})  & \ang{;;4.23} $\times$ \ang{;;3.35} (\ang{-10.3}) \\
Per-emb 22  & 29 Nov 2014 & 22 Sep 2015 & 152 &  1.96 & \ang{;;3.77} $\times$ \ang{;;2.98} (\ang{-15.1}) & \ang{;;4.21} $\times$ \ang{;;3.12} (\ang{-25.5}) \\
Per-emb 25  & 26 Oct 2015 & ...         & 175 &  2.61 & \ang{;;4.11} $\times$ \ang{;;2.98} (\ang{39.2})  & \ang{;;3.43} $\times$ \ang{;;2.98} (\ang{63.5})  \\
Per-emb 26/42 & 18 Nov 2014 & ...         & 112 &  1.34 & \ang{;;4.38} $\times$ \ang{;;3.38} (\ang{-20.3}) & \ang{;;4.06} $\times$ \ang{;;3.26} (\ang{-12.1}) \\
Per-emb 27  & 20 Nov 2014 & 04 Sep 2014 & 114 &  1.97 & \ang{;;2.34} $\times$ \ang{;;2.22} (\ang{41.5})  & \ang{;;4.14} $\times$ \ang{;;3.26} (\ang{-12.7}) \\
Per-emb 29  & 28 Nov 2014 & 22 Sep 2015 & 132 &  2.05 & \ang{;;3.59} $\times$ \ang{;;2.87} (\ang{4.3})   & \ang{;;4.31} $\times$ \ang{;;3.05} (\ang{-18.9}) \\
Per-emb 33  & 18 Nov 2014 & ...         & 227 &  3.57 & \ang{;;4.35} $\times$ \ang{;;3.36} (\ang{-19.7}) & \ang{;;4.04} $\times$ \ang{;;3.25} (\ang{-12.0}) \\
Per-emb 35  & 13 Dec 2014 & 06 Oct 2015 &  85 &  1.74 & \ang{;;3.17} $\times$ \ang{;;2.77} (\ang{5.9})   & \ang{;;4.07} $\times$ \ang{;;3.58} (\ang{8.7})   \\
Per-emb 44  & 19 Oct 2015 & ...         & 225 &  2.85 & \ang{;;3.50} $\times$ \ang{;;3.13} (\ang{78.9})  & \ang{;;3.42} $\times$ \ang{;;3.05} (\ang{82.0})  \\
Per-emb 47  & 19 Oct 2015 & ...         &  98 &  1.55 & \ang{;;3.50} $\times$ \ang{;;3.15} (\ang{76.3})  & \ang{;;3.42} $\times$ \ang{;;3.02} (\ang{88.1})  \\
Per-emb 53  & 30 Nov 2014 & ...         & 179 &  2.09 & \ang{;;4.58} $\times$ \ang{;;3.37} (\ang{-10.1}) & \ang{;;4.38} $\times$ \ang{;;3.46} (\ang{-9.7})  \\
Per-emb 61  & 30 Nov 2014 & ...         & 171 &  1.93 & \ang{;;4.52} $\times$ \ang{;;3.36} (\ang{-11.8}) & \ang{;;4.36} $\times$ \ang{;;3.43} (\ang{-11.6}) \\
\hline
\end{tabular}
\tablefoot{
\tablefoottext{a}{Date of observations. If both subcompact and extended data are available, they are concatenated.}
\tablefoottext{b}{1$\upsigma$ RMS noise of the \SI{1.3}{mm} continuum and the C$^{18}$O\,2--1 moment zero maps. The units are respectively \si{mJy.beam^{-1}} and \si{mJy.beam^{-1}.km.s^{-1}}.}
\tablefoottext{c}{Synthesised beam sizes. The beam's position angle is measured relative to the major axis from North through East.}
}
\end{table*}

One way of using chemistry to trace protostellar accretion is to measure the size of the emitting region of CO or some other molecule towards individual protostars. If the emission is detected over larger spatial scales than can be explained by the current protostellar luminosity, it will indicate that the luminosity has been larger in the past, and that the molecule is still in the process of refreezing. Such an approach was adopted by \citet{Jorgensen:2015kz}, who measured the sizes of the C$^{18}$O\,$J$\,=\,2--1 emitting regions towards a sample of 16 deeply embedded protostars and found half to show evidence of a past accretion burst. Using a large three-dimensional MHD simulation of a molecular cloud, \citet{Frimann:2016ke} calculated synthetic \co maps of a large sample of simulated protostellar systems and measured the sizes of the \co emitting regions in the same way as was done by \citeauthor{Jorgensen:2015kz}. The study confirmed that the approach taken by \citeauthor{Jorgensen:2015kz} is a valid one since it demonstrated that the sizes of the \co emission regions can be accurately measured towards systems with realistic core morphologies by an interferometer with a baseline coverage limited to ranges between 15 and 50\,k$\uplambda$ (corresponding to angular scales between roughly \ang{;;6} and \ang{;;2}).

\begin{figure*}
\includegraphics[width=18cm]{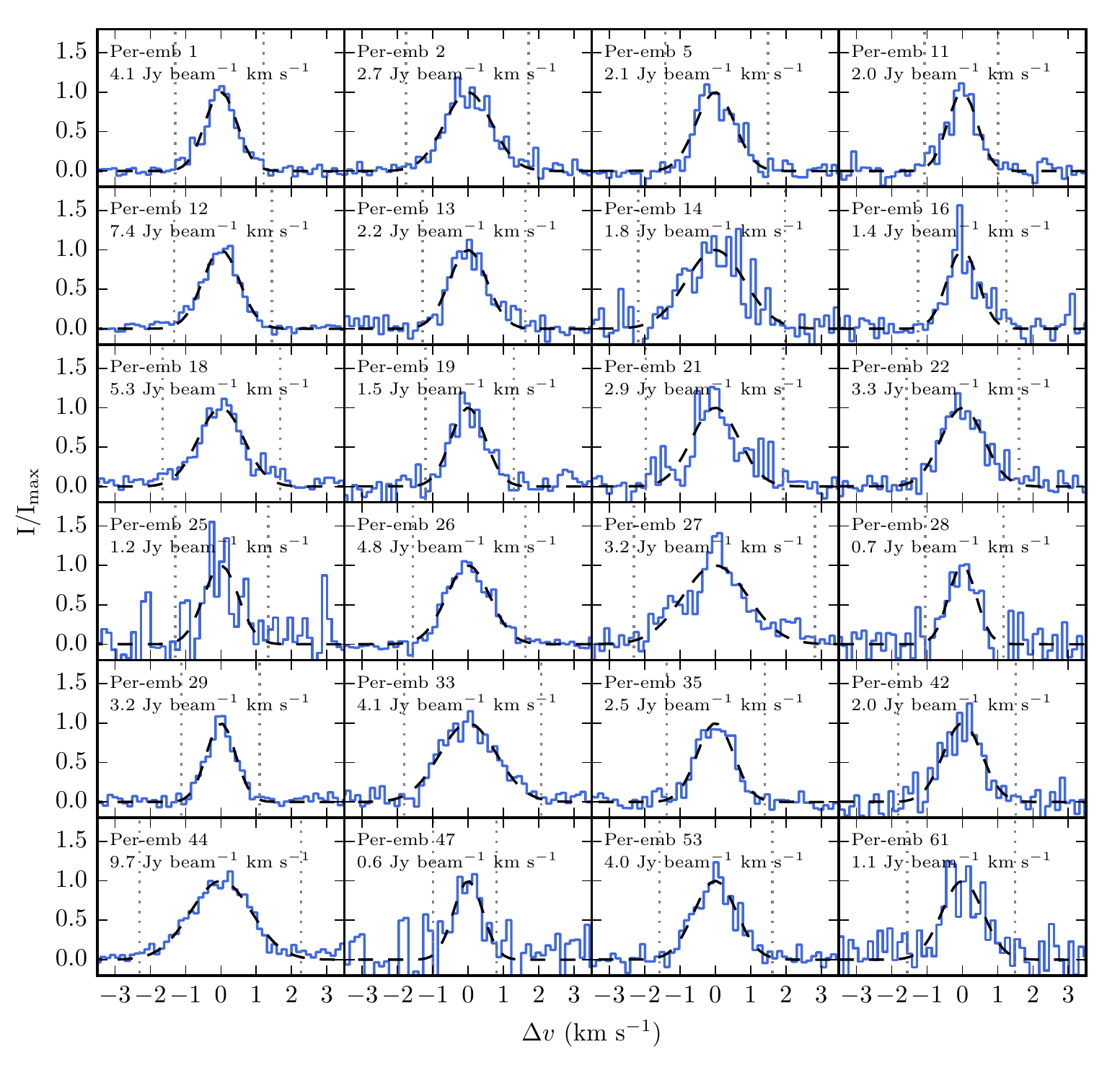}
   \caption{C$^{18}$O~2--1 spectra towards the peak of the integrated emission of the observed sources. The integration intervals used for producing the moment zero maps in Fig.~\ref{fig:c18omap} are indicated by vertical dotted lines and run from $-2.5\upsigma$ to $2.5\upsigma$ of the fitted Gaussians (black dashed lines). The numeric value in the top left corner of each panel is the integrated emission over the line.}
   \label{fig:spectrum}
\end{figure*}

This paper presents continuum and C$^{18}$O~2--1 observations towards a sample of 24 embedded protostars in the Perseus molecular cloud (distance: \SI{235}{pc}; \citealt{Hirota:2008gu}). The aim of the paper is to measure the sizes of the \co emitting regions towards the observed protostars and to use those measurements as a tracer of past accretion bursts. Furthermore, the continuum observations are used to investigate whether a link between accretion bursts and circumstellar disks can be established. The key advantages of this sample, relative to the one analysed in \citet{Jorgensen:2015kz}, are that all the protostars are from the same molecular cloud, meaning that distance uncertainties do not play a role, and that the star formation environments are similar. Also, the observations are all from the same program, which provides a more uniform sample. The observations are from the large survey ``Mass Assembly of Stellar Systems and their Evolution with the SMA'' (MASSES; Co-PIs: Michael M.\ Dunham and Ian W.\ Stephens) undertaken with the Submillimeter Array (SMA). The survey targets all known embedded protostars in Perseus, including the \num{66} sources identified with the Spitzer Space Telescope by \citet{Enoch:2009ch} and seven candidate first hydrostatic cores. The survey is still ongoing, hence this paper comprises the sample of sources imaged thus far. First results from MASSES were presented by \citet{Lee:2015il,Lee:2016fh}.

The paper is laid out as follows: Section~\ref{sec:observations} describes the observations and introduces the observed sample. Section~\ref{sec:fitting} describes the procedure used for measuring the sizes of the \co emitting regions, while Sect.~\ref{sec:results} presents the results of the analysis of both the \co and continuum data. Depending on the adopted CO sublimation temperature, between \SI{20}{\percent} and \SI{50}{\percent} of the observed sources show evidence of a past accretion burst, which is consistent with the results of \citet{Jorgensen:2015kz}. Section~\ref{sec:discussion} presents a discussion of the results including possible links between episodic accretion and circumstellar disks and between episodic accretion and multiple systems. Finally, Sect.~\ref{sec:Summary} summarises the findings of the paper.

\section{Observations}
\label{sec:observations}

\subsection{SMA}
The observations were carried out with the SMA \citep{Ho:2004bz}, a submillimetre and millimetre interferometer, consisting of eight \SI{6}{m} antennae situated on the summit of Mauna Kea, Hawaii. The data include observations of the \co~2--1 transition (rest frequency: \SI{219.56}{GHz}) and \SI{230}{GHz} (\SI{1.3}{mm}) dust continuum observations. All of the data were taken in the array's subcompact configuration, which typically covers baselines between 5\,k$\uplambda$ and 50\,k$\uplambda$. For some targets, C$^{18}$O observations from the extended configuration are also available (typical baseline coverage between 20\,k$\uplambda$ and 150\,k$\uplambda$), which are concatenated with the subcompact data whenever available. The data were reduced and calibrated with the MIR package\footnote{\url{https://www.cfa.harvard.edu/~cqi/mircook.html}} using standard calibration procedures, while imaging was done using Miriad \citep{Sault:1995ub}. Table~\ref{tbl:object} lists the objects in the observed sample along with their Spitzer positions, luminosities, and bolometric temperatures, while Table~\ref{tbl:obs2} presents a log of the observations. For further details on the data reduction process, the reader is referred to \citet{Lee:2015il}.

\begin{figure*}
\includegraphics[width=18cm]{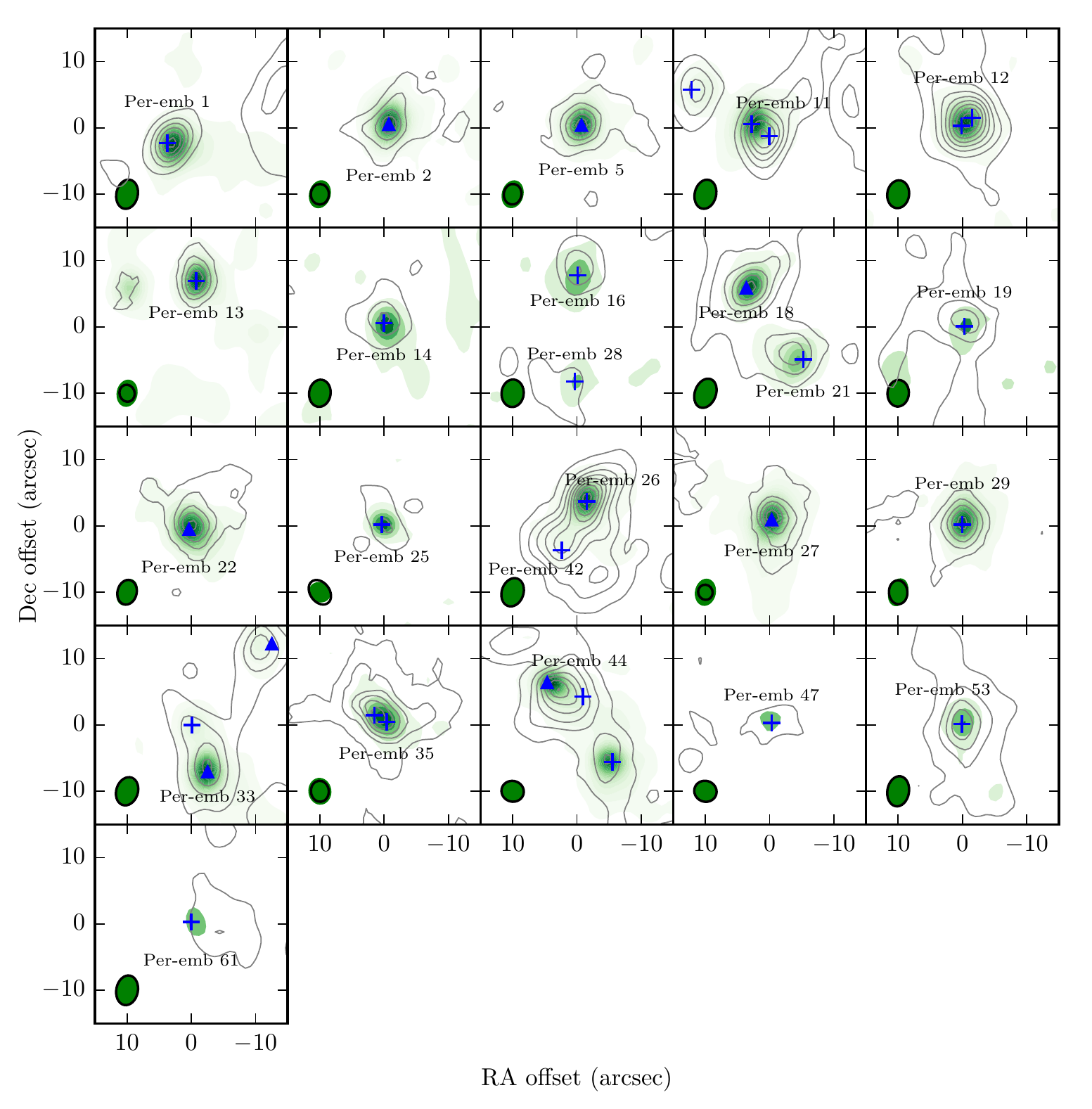}
   \caption{Maps of \co~2--1 integrated emission (line contours) and \SI{1.3}{mm} continuum emission (colour map). Contours are drawn starting from 3$\upsigma$ in steps of 6$\upsigma$ (see Table~\ref{tbl:obs2} for \co and continuum 1$\upsigma$ RMS sensitivities). Synthesised beams of the C$^{18}$O and continuum emission are shown in the lower left corners as respectively black and a shaded ellipses. Note that the beam sizes only differ significantly for the targets where extended data is available for the \co emission. The blue symbols mark the source positions from the VANDAM survey \citep{Tobin:2016fl}, with crosses indicating single protostars and triangles indicating multiple systems with projected separations $<$\ang{;;0.8}.}
   \label{fig:c18omap}
\end{figure*}

Figure~\ref{fig:spectrum} shows \co~2--1 spectra extracted towards the peak of the integrated emission of each source. The area between the dotted lines indicate the integration intervals used to produce the C$^{18}$O moment zero maps shown in Fig.~\ref{fig:c18omap}. The integration intervals are determined by Gaussian fits and run between $-2.5\upsigma$ and $2.5\upsigma$. As seen from the spectra, essentially all of the emission is recovered towards the individual sources. Only Per-emb~27 has its upper integration limit increased by \SI{0.5}{km.s^{-1}} to catch additional red-shifted emission. Figure~\ref{fig:c18omap} also shows continuum maps of the observed sources. The emission is observed to be relatively compact towards the individual sources, with some evidence of low surface brightness extended emission toward some sources.

The blue symbols in Fig.~\ref{fig:c18omap} indicate source positions from the ``VLA Nascent Disk and Multiplicity Survey of Perseus Protostars'' (VANDAM) survey \citep{Tobin:2016fl}. Crosses are used for single protostars, while triangles indicate multiple systems with projected separations $<$\ang{;;0.8}. The triangles all represent binary systems except for Per-emb~33, which is a triple system. Of the 24 sources in the sample, ten are part of multiple systems not resolved in the SMA maps. The sources are labelled according to the naming scheme of \citet{Enoch:2009ch}, who used Spitzer observations to identify the protostars. The size of the Spitzer beam at \SI{24}{\micro\metre} is \ang{;;7}, meaning that if a multiple system is not resolved in the SMA observations it will not have been resolved in the Spitzer \SI{24}{\micro\metre} observations either, and the label covers all members of the system.

\begin{figure*}
\centering
\includegraphics[width=18cm]{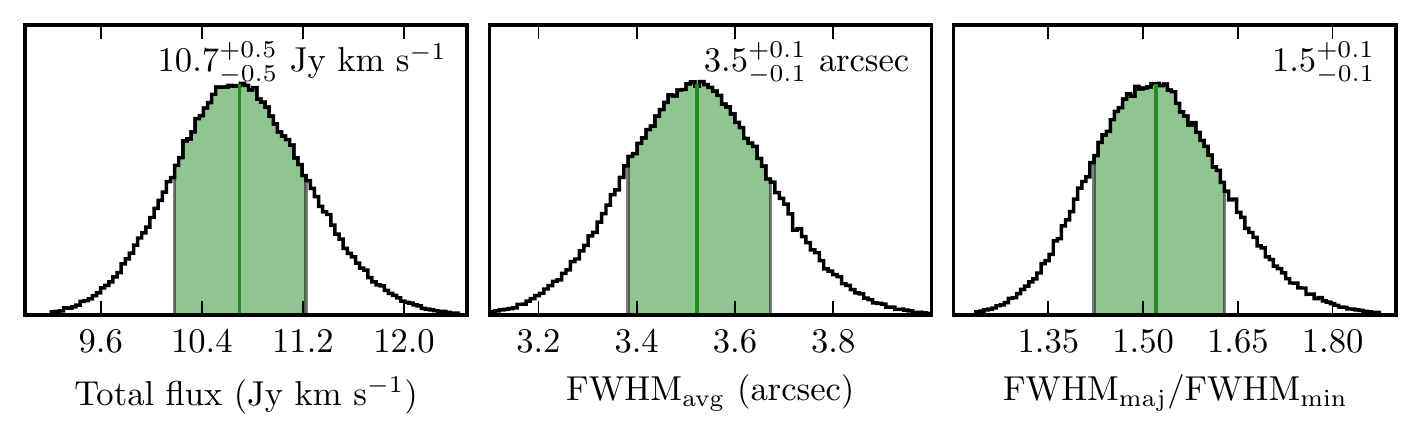}
   \caption{Examples of posterior parameter distributions of three fitted parameters towards Per-emb~1. The green lines represent the medians of the distributions and the shaded areas the 1$\upsigma$ uncertainty region. The median value and uncertainty are printed in the upper right corner of each panel.}
   \label{fig:dist}
\end{figure*}

The addition of extended SMA data does not alter the \co maps. This is related to the fact that \co is typically not detected at baselines $\gtrsim$\,50\,k$\uplambda$. The addition of extended data is still advantageous, as it provides additional baselines down to $\approx$20\,k$\uplambda$, thereby improving the signal at intermediate baselines. As a test of consistency between the samples, the \co~2--1 observations of the nine Perseus protostars from the \citet{Jorgensen:2015kz} survey are also included. Only one of these nine sources, IRAS~03256, is not part of the sample analysed here. Going forward, the two samples are distinguished by referring to the `MASSES' sample and the `\jor' sample respectively.

\subsection{JCMT/SCUBA}

The large-scale emission of the observed sources is traced with \SI{850}{\micro\metre} single-dish continuum observations from the James Clerk Maxwell Telescope Submillimetre-Common Bolometer Array (JCMT/SCUBA) legacy catalogue \citep{DiFrancesco:2008di}. The SCUBA maps have a resolution of \ang{;;15}, and all the sources in the sample, except for Per-emb~47, are detected.

\section{Model fitting}
\label{sec:fitting}

To measure the sizes of the \co emitting regions towards the observed sources, we follow the approach of \citet{Jorgensen:2015kz} and \citet{Frimann:2016ke} and fit Gaussians to the interferometric data directly in the $(u,v)$-plane, thereby avoiding uncertainties introduced by the deconvolution procedure. Miriad's \texttt{uvfit} routine can be used to fit the Gaussians, but we find that its results can be sensitive to the starting guess of the fitting parameters, which indicates that it does not always converge to the optimal result. To circumvent this issue, the Gaussians are instead fitted using the Python Markov Chain Monte Carlo (MCMC) code \texttt{emcee}\footnote{\url{http://dan.iel.fm/emcee/current/}} \citep{ForemanMackey:2013io}. An MCMC method has the advantage that it is capable of exploring the parameter space better relative to a standard non-linear fitting algorithm, like the one used by \texttt{uvfit}.

Any MCMC algorithm needs a likelihood function describing the probability of the data given a set of input parameters, as well as prior probabilities of the parameters. The likelihood function used here takes the form
\begin{equation}
  \mathcal{L}\left(\mathbf{p}\right) = \left(\sum_{i=1}^N \left[d_i - m_i\left(\mathbf{p}\right)\right]^2\right)^{-\frac{N}{2}}, \nonumber
\end{equation}
where the sum is over the individual data points, $d_i$, and $m\left(\mathbf{p}\right)$ is the model function with fitting parameters $\mathbf{p}$. This likelihood function is appropriate for situations where one is ignorant about the uncertainties on the individual data points (e.g.\ Chapter~9 of \citealt{Gregory:2005aa}). The model function, $m$, is a two-dimensional Gaussian function described by six parameters: the total flux; the peak position relative to the phase-centre in the RA and Dec directions; the average FWHM of the Gaussian, $\mathrm{FWHM}_\mathrm{avg} = \sqrt{\mathrm{FWHM}_\mathrm{min} \times \mathrm{FWHM}_\mathrm{maj}}$, where $\mathrm{FWHM}_\mathrm{min}$ and $\mathrm{FWHM}_\mathrm{maj}$ are the minor and major axes; the ratio between the major and minor axes, $\mathrm{FWHM}_\mathrm{maj}/\mathrm{FWHM}_\mathrm{min}$; and the position angle of the Gaussian. We are largely ignorant about the prior probabilities and therefore set them as widely as possible. For the total flux, $\mathrm{FWHM}_\mathrm{avg}$, and $\mathrm{FWHM}_\mathrm{maj}/\mathrm{FWHM}_\mathrm{min}$ the prior probabilities are assumed to be flat in log-space between \SIrange[range-phrase={ and }]{0.1}{100}{Jy.km.s^{-1}}, \ang{;;0.1} and~\ang{;;100}, and 1 and~100. For the peak position the prior probability is assumed to be flat within a radius of \ang{;;5} from the peak position estimated from the moment zero maps. Finally, the prior probability of the position angle is assumed to be flat between \ang{0} and~\ang{180}. We note that as long as the priors are set wide enough, they do not influence the fitting results.

The result of the MCMC fitting is the posterior probability distribution, which can be marginalised and drawn as one-dimensional distributions for each parameter (see Fig.~\ref{fig:dist} for examples). The distributions are typically, but not always, well represented by Gaussians, and the results are therefore reported by giving the distribution's median with the 16th and 84th percentiles approximating the lower and upper bound of the uncertainty. The sources are also fitted using Miriad's \texttt{uvfit} routine. The results of the two methods deviate significantly for five out of 24 sources. For those sources where the two methods do not agree, we confirm that \texttt{emcee} converges to a result with a lower sum of the squared residuals than \texttt{uvfit}.

\section{Results}
\label{sec:results}

\begin{figure*}
\centering
\includegraphics[width=18cm]{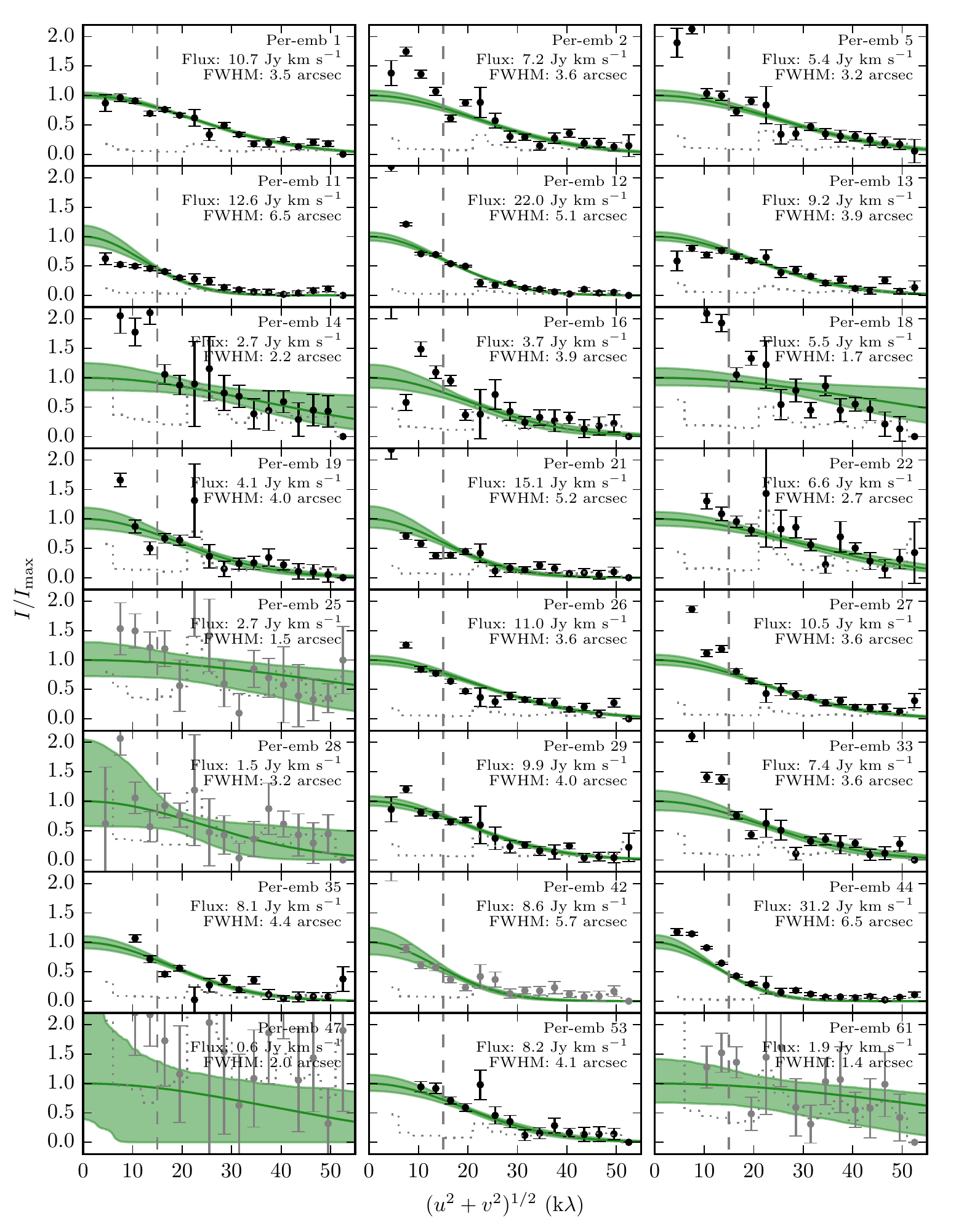}
   \caption{$(u,v)$-amplitude plots of the integrated \co emission towards the sources in the observed sample. The grey dotted histograms in each panel indicate the expected amplitude in the absence of any source emission. The solid green line in each panel represents the best fitting Gaussian, while the shaded region indicate the 1$\upsigma$ uncertainty region of the fit. The vertical axis of each panel has been scaled to the Gaussian's peak flux, whose value is printed in the upper right. The grey dashed lines indicate the 15\,k$\uplambda$ lower boundary of the fit. Panels where the $(u,v)$-amplitudes have been plotted in grey indicate objects whose fits have been rejected (see text for details).}
   \label{fig:ampaccept}
\end{figure*}

\subsection{C$^{18}$O observations}
\label{sec:c18oobs}

\begin{table*}
\caption{Gaussian fitting results towards the integrated \co emission.}
\label{tbl:results}
\centering
\begin{tabular}{l S[separate-uncertainty,table-format=1.2(2)] S[table-format=1.1]@{\hskip 2pt}l S[table-format=1.2]@{\hskip 2pt}l S[table-format=2.2]@{\hskip 2pt}l S[table-format=3.1]@{\hskip 2pt}l S[table-format=1.1]@{\hskip 2pt}l}
\hline\hline
Source & {$L_\mathrm{bol}$} & \multicolumn{2}{c}{$\mathrm{FWHM}_\mathrm{avg}$} & \multicolumn{2}{c}{$\mathrm{FWHM}_\mathrm{major}/\mathrm{FWHM}_\mathrm{minor}$} & \multicolumn{2}{c}{$\left(L_\mathrm{CO}/L_\mathrm{cur}\right)_{21\,\mathrm{K}}$\tablefootmark{a}} & \multicolumn{2}{c}{$\left(L_\mathrm{CO}/L_\mathrm{cur}\right)_{28\,\mathrm{K}}$\tablefootmark{a}} & \multicolumn{2}{c}{$\mathrm{FWHM}_\mathrm{avg,JVWB15}$\tablefootmark{b}} \\
 & {($L_\sun$)} & \multicolumn{2}{c}{(arcsec)} & {} & {} & {} & {} & {} & {} & \multicolumn{2}{c}{(arcsec)}\\
\hline
Per-emb 1  & 1.8(1)  &  3.5 & $\substack{+0.1 \\ -0.1}$    &  1.5 & $\substack{+0.1 \\ -0.1}$    &  3.3 & $\substack{+0.4 \\ -0.3}$    &   16 & $\substack{+2 \\ -2}$        & {...}                               \\[2pt]
Per-emb 2  & 0.90(7) &  3.6 & $\substack{+0.2 \\ -0.2}$    &  1.4 & $\substack{+0.2 \\ -0.1}$    &    7 & $\substack{+1 \\ -1}$        &   33 & $\substack{+7 \\ -6}$        & {...}                               \\[2pt]
Per-emb 5  & 1.3(1)  &  3.2 & $\substack{+0.2 \\ -0.2}$    &  1.2 & $\substack{+0.1 \\ -0.1}$    &  3.5 & $\substack{+0.7 \\ -0.6}$    &   17 & $\substack{+3 \\ -3}$        &  3.4 & $\substack{+0.1 \\ -0.1}$    \\[2pt]
Per-emb 11 & 1.5(1)  &  6.5 & $\substack{+0.5 \\ -0.4}$    &  1.6 & $\substack{+0.1 \\ -0.1}$    &   16 & $\substack{+3 \\ -3}$        &   80 & $\substack{+20 \\ -10}$      & {...}                               \\[2pt]
Per-emb 12 & 7.0(7)  &  5.1 & $\substack{+0.2 \\ -0.2}$    & 1.06 & $\substack{+0.06 \\ -0.04}$  &  2.0 & $\substack{+0.3 \\ -0.3}$    &   10 & $\substack{+2 \\ -1}$        &  5.3 & $\substack{+0.5 \\ -0.4}$    \\[2pt]
Per-emb 13 & 4.0(3)  &  3.9 & $\substack{+0.2 \\ -0.2}$    &  1.8 & $\substack{+0.2 \\ -0.1}$    &  1.9 & $\substack{+0.3 \\ -0.3}$    &    9 & $\substack{+2 \\ -1}$        &  0.5 & $\substack{+0.6 \\ -0.3}$    \\[2pt]
Per-emb 14 & 0.70(8) &  2.2 & $\substack{+0.8 \\ -1.5}$    &  1.9 & $\substack{+13.8 \\ -0.7}$   &    3 & $\substack{+3 \\ -3}$        &   10 & $\substack{+20 \\ -10}$      & {...}                               \\[2pt]
Per-emb 16 & 0.40(4) &  3.9 & $\substack{+0.8 \\ -0.5}$    &  1.2 & $\substack{+0.2 \\ -0.2}$    &   19 & $\substack{+10 \\ -6}$       &   90 & $\substack{+50 \\ -30}$      & {...}                               \\[2pt]
Per-emb 18 & 3.6(5)  &  1.7 & $\substack{+0.8 \\ -1.1}$    &    4 & $\substack{+26 \\ -2}$       &  0.3 & $\substack{+0.5 \\ -0.3}$    &    1 & $\substack{+2 \\ -1}$        & {...}                               \\[2pt]
Per-emb 19 & 0.36(5) &  4.0 & $\substack{+0.5 \\ -0.5}$    &  1.5 & $\substack{+0.3 \\ -0.2}$    &   23 & $\substack{+9 \\ -6}$        &  110 & $\substack{+50 \\ -30}$      & {...}                               \\[2pt]
Per-emb 21 & 3.6(5)  &  5.2 & $\substack{+0.6 \\ -0.5}$    &  1.5 & $\substack{+0.2 \\ -0.2}$    &    4 & $\substack{+1 \\ -1}$        &   21 & $\substack{+8 \\ -5}$        & {...}                               \\[2pt]
Per-emb 22 & 3.6(5)  &  2.7 & $\substack{+0.3 \\ -0.4}$    &  1.6 & $\substack{+0.4 \\ -0.2}$    &  0.9 & $\substack{+0.3 \\ -0.3}$    &    4 & $\substack{+2 \\ -1}$        &  4.0 & $\substack{+0.5 \\ -0.5}$    \\[2pt]
Per-emb 25\tablefootmark{c} & 0.95(2) &  1.5 & $\substack{+1.4 \\ -0.8}$    &   20 & $\substack{+40 \\ -10}$      &  0.8 & $\substack{+3.2 \\ -0.7}$    &    4 & $\substack{+15 \\ -3}$       & {...}                               \\[2pt]
Per-emb 26 & 9(1)    &  3.6 & $\substack{+0.2 \\ -0.2}$    &  1.3 & $\substack{+0.1 \\ -0.1}$    &  0.7 & $\substack{+0.1 \\ -0.1}$    &  3.4 & $\substack{+0.7 \\ -0.6}$    &  3.4 & $\substack{+0.9 \\ -1.0}$    \\[2pt]
Per-emb 27 & 19.0(4) &  3.6 & $\substack{+0.3 \\ -0.2}$    &  1.4 & $\substack{+0.1 \\ -0.1}$    & 0.33 & $\substack{+0.06 \\ -0.05}$  &  1.6 & $\substack{+0.3 \\ -0.2}$    &    5 & $\substack{+1 \\ -1}$        \\[2pt]
Per-emb 28\tablefootmark{c} & 0.70(8) &    3 & $\substack{+3 \\ -2}$        &    3 & $\substack{+25 \\ -1}$       &    7 & $\substack{+24 \\ -7}$       &   30 & $\substack{+130 \\ -30}$     & {...}                               \\[2pt]
Per-emb 29 & 3.7(4)  &  4.0 & $\substack{+0.3 \\ -0.2}$    & 1.13 & $\substack{+0.13 \\ -0.09}$  &  2.1 & $\substack{+0.4 \\ -0.3}$    &   11 & $\substack{+2 \\ -2}$        & {...}                               \\[2pt]
Per-emb 33 & 8.3(8)  &  3.6 & $\substack{+0.5 \\ -0.5}$    &  2.5 & $\substack{+0.5 \\ -0.4}$    &  0.8 & $\substack{+0.3 \\ -0.2}$    &    4 & $\substack{+2 \\ -1}$        &    5 & $\substack{+3 \\ -1}$        \\[2pt]
Per-emb 35 & 9.1(3)  &  4.4 & $\substack{+0.3 \\ -0.3}$    &  2.2 & $\substack{+0.2 \\ -0.2}$    &  1.1 & $\substack{+0.2 \\ -0.2}$    &  5.4 & $\substack{+1.0 \\ -0.9}$    & {...}                               \\[2pt]
Per-emb 42\tablefootmark{c} & 0.7(8)  &  5.7 & $\substack{+0.6 \\ -0.6}$    &  1.5 & $\substack{+0.2 \\ -0.2}$    &   20 & $\substack{+37 \\ -10}$      &  100 & $\substack{+190 \\ -50}$     & {...}                               \\[2pt]
Per-emb 44 & 32(7)   &  6.5 & $\substack{+0.5 \\ -0.4}$    & 1.28 & $\substack{+0.07 \\ -0.06}$  &  0.8 & $\substack{+0.3 \\ -0.2}$    &  4.0 & $\substack{+1.4 \\ -0.9}$    &  6.4 & $\substack{+0.2 \\ -0.2}$    \\[2pt]
Per-emb 47\tablefootmark{c} & 1.2(1)  &    2 & $\substack{+28 \\ -2}$       &   11 & $\substack{+37 \\ -8}$       &    1 & $\substack{+706 \\ -1}$      &    6 & $\substack{+4359 \\ -6}$     & {...}                               \\[2pt]
Per-emb 53 & 4.7(9)  &  4.1 & $\substack{+0.4 \\ -0.4}$    & 1.11 & $\substack{+0.14 \\ -0.08}$  &  1.9 & $\substack{+0.7 \\ -0.5}$    &    9 & $\substack{+4 \\ -2}$        & {...}                               \\[2pt]
Per-emb 61\tablefootmark{c} & 0.2(2)  &  1.4 & $\substack{+1.6 \\ -0.8}$    &   12 & $\substack{+38 \\ -9}$       &    3 & $\substack{+17 \\ -3}$       &   10 & $\substack{+80 \\ -10}$      & {...}                               \\[2pt]
\hline
Per-emb 2  & 0.90(7) &  5.0 & $\substack{+0.3 \\ -0.3}$    &  1.8 & $\substack{+0.2 \\ -0.1}$    &   15 & $\substack{+3 \\ -2}$        &   70 & $\substack{+10 \\ -10}$      & {...}                               \\[2pt]
Per-emb 16 & 0.40(4) &  7.6 & $\substack{+0.7 \\ -0.7}$    &  1.5 & $\substack{+0.1 \\ -0.1}$    &   90 & $\substack{+20 \\ -20}$      &  500 & $\substack{+100 \\ -100}$    & {...}                               \\[2pt]
Per-emb 18 & 3.6(5)  &  5.8 & $\substack{+0.9 \\ -1.0}$    &  2.1 & $\substack{+0.4 \\ -0.4}$    &    5 & $\substack{+2 \\ -2}$        &   30 & $\substack{+10 \\ -10}$      & {...}                               \\[2pt]
\hline
\end{tabular}
\tablefoot{Measurements in the top part of the table are performed using a lower baseline limit of 15\,k$\uplambda$. Lowering the limit to 10\,k$\uplambda$ changes the results substantially for Per-emb~2, 16, and~18, and those results are listed below the horizontal line.
\tablefoottext{a}{$L_\mathrm{CO}$ is calculated by solving for the luminosity and inserting the measured $\mathrm{FWHM}_\mathrm{avg}$ in Eq.~\eqref{eq:extent}. The subscript indicate whether a CO sublimation temperature of \SI{21}{K} or \SI{28}{K} has been adopted.}
\tablefoottext{b}{\jor measurements.}
\tablefoottext{c}{Excluded sources.}
}
\end{table*}

Using the method described in Sect.~\ref{sec:fitting}, two-dimensional Gaussians are fitted to the integrated C$^{18}$O visibilities of all 24 sources in the sample. The smallest baselines, which correspond to large-scale emission, are excluded as they may include emission that is not regulated by the protostellar heating. Emission from extended spatial scales may instead be the result of heating by the interstellar radiation field, or it may originate from low-density pre-depletion regions where CO has not yet frozen out \citep{Jorgensen:2005cs}. Here, we follow \citet{Jorgensen:2015kz} and only include baselines above 15\,k$\uplambda$, corresponding to angular scales $\lesssim$\,\ang{;;6}.

Figure~\ref{fig:ampaccept} shows $(u,v)$-amplitudes and Gaussian fits of the observed sources. The $(u,v)$-amplitudes implicitly assume axial symmetry around the centre so, for those fields that contain multiple emission peaks, the emission that does not pertain to the source of interest is subtracted out before plotting the $(u,v)$-amplitudes. The subtraction is done by fitting a Gaussian to the secondary emission in the moment zero map and then subtracting that Gaussian directly from the interferometric visibilities; if necessary this procedure may be repeated several times. The subtraction is solely for the benefit of being able to display the $(u,v)$-amplitudes of sources located in fields with multiple emission peaks. For the actual model fitting, the original non-subtracted data are always used, and multiple Gaussians are fitted to fields with more than one emission peak.

Table~\ref{tbl:results} shows the results of the Gaussian fits to the sources in the MASSES sample. For most of the sources, the statistical uncertainties of the measured values of $\mathrm{FWHM}_\mathrm{avg}$ (hereafter referred to as the measured extents) are less than \ang{;;0.5}. Also, the fitted Gaussians are relatively circular, with most fits having $\mathrm{FWHM}_\mathrm{major}/\mathrm{FWHM}_\mathrm{minor}$\,$\leq$\,1.5 (see Sect.~\ref{sec:asymmetry} for a discussion of the causes of asymmetric emission).

Per-emb~25, 28, 47, and~61 are excluded from further analysis, as their signal-to-noise ratios are too small to measure the extents reliably. The low signal-to-noise ratios may themselves be an indirect clue that the four sources have not recently undergone an accretion burst, since the elevation of CO ice into the gas phase is expected to enhance the strength of the emission (\citealt{Visser:2015ew}; also see discussion in Sect.~\ref{sec:reliability}). Three of the sources have a measured $L_\mathrm{bol}$<\,$1\,L_\sun$, and it is therefore not surprising that their emission is too weak to be measured reliably. The low signal-to-noise ratios may also indicate that the sources are surrounded by small amounts of envelope material. This scenario seems particularly likely for Per-emb~47, which is not detected in the SCUBA maps, but is also likely for both Per-emb~25 and~63, both of which have small inferred envelope masses (cf.\ Sect.~\ref{sec:continuum}). Naturally, the two scenarios are not mutually exclusive. Per-emb~42 is excluded from the subsequent analysis, not because of a lack of signal, but because its emission cannot be properly untangled from the emission of the nearby 9.2\,$L_\sun$ protostar Per-emb~26.

\begin{figure}
\includegraphics[width=\hsize]{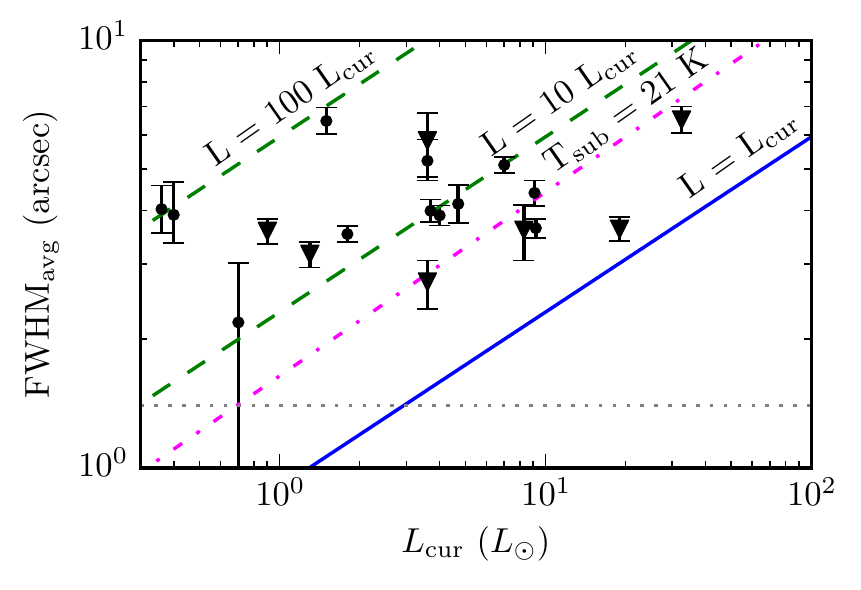}
\caption{Measured C$^{18}$O extents vs.\ the current (bolometric) luminosity. Triangular symbols represent sources that are associated with a companion protostar at a projected distance <\ang{;;0.8}. The error-bars indicate the statistical uncertainties of the measurements listed in Table~\ref{tbl:results}. The uncertainties of the luminosities are not shown to avoid cluttering the diagram, but they are small enough that they do not influence the conclusions. The solid line represents the expected CO extent given the current luminosity and a CO sublimation temperature of \SI{28}{K}, while the dashed lines represent the expected CO extents for 10 and 100 times the current luminosity. The dash-dot line shows the expected CO extent for a sublimation temperature of \SI{21}{K}. The horizontal dotted line at \ang{;;1.4} indicates the lower limit of the extents that can be measured given the baseline coverage and sensitivity of the observations.}
\label{fig:extent}
\end{figure}

Figure~\ref{fig:extent} shows the measured \co extents vs.\ the current (bolometric) luminosity, $L_\mathrm{cur}$, of the protostars in the MASSES sample. In the absence of any past accretion variability and for an adopted CO sublimation temperature of \SI{28}{K}, the measurements are expected to follow the blue line, which has been calculated from synthetic observables of a set of axissymmetric radiative transfer models, presented in Appendix~\ref{sec:appB}. The line is described by the equation
\begin{equation}
  \mathrm{FWHM}_\mathrm{avg} = a \left(\frac{d}{\SI{235}{pc}}\right)^{-1} \left(\frac{L_\mathrm{bol}}{1\,L_\sun}\right)^{0.41},
  \label{eq:extent}
\end{equation} 
where $d$ is the distance to the source, and the angular coefficient, $a$, depends on the adopted sublimation temperature. For $T_\mathrm{sub} = \SI{28}{K}$, appropriate for CO ice mixed with water \citep{Noble:2012hq}, $a = \ang{;;0.89}$. For $T_\mathrm{sub} = \SI{21}{K}$, appropriate for pure CO ice \citep{Sandford:1993da,Bisschop:2006dc}, $a = \ang{;;1.64}$. 

Many of the sources shown in Fig.~\ref{fig:extent} have measured extents that are larger than predicted, particularly for the sources with $L_\mathrm{bol} \lesssim 2\,L_\sun$. It is natural to suspect that this could be an indication of an instrumental bias of the interferometer. We will argue in Sect.~\ref{sec:reliability} that this is not the case, and that the measured extents are reliable. However, at the same time it is also likely that the four sources that were excluded due to too small signal-to-noise ratios have small C$^{18}$O emitting regions that could not be accurately measured due to a combination of sensitivity and baseline issues.

Choosing a lower baseline limit of 15\,k$\uplambda$ has the effect of excluding all emission on angular scales $\gtrsim$\,\ang{;;6}. Such a limit clearly inhibits the measurement of intense bursts that elevate CO into the gas phase over large regions, but it is necessary to avoid contamination from regions where the C$^{18}$O emission is not regulated by the heating from the central protostar. Based on measurements of synthetic data, \citet{Frimann:2016ke} found that it is typically safe to include baselines down to 10\,k$\uplambda$ (angular scale of \ang{;;9}), before one begins to run into issues with large-scale emission not associated with the heating from the central source. However, it should be noted that the synthetic data did not include sources with multiple emission peaks, which means that the large-scale emission was less confused than is often the case here. Redoing the analysis with a lower boundary of 10\,k$\uplambda$ alters the results substantially (by more than 1$\upsigma$ in the fitting uncertainty) for only three objects, Per-emb~2, 16, and~18, that are listed in the bottom of Table~\ref{tbl:results}. For Per-emb~2, it is difficult to judge which of the two measurements is most accurate, but both yield a measured extent that is significantly enhanced relative to what is expected from the current luminosity. Since the difference is not crucial to the conclusions, we retain the lower boundary of 15\,k$\uplambda$, which gives the most conservative estimate of the burst magnitude. For Per-emb~16, the measured extent, given a baseline limit of 10\,k$\uplambda$, is significantly larger than the apparent extent of the source in the moment zero map, and we therefore retain the 15\,k$\uplambda$ limit. For Per-emb~18, it can be seen in the moment zero map that the C$^{18}$O emission towards the source is indeed very extended, and the 10\,k$\uplambda$ measurement is adopted going forward.

By solving Eq.~\eqref{eq:extent} for the luminosity, it is possible to calculate the luminosity, $L_\mathrm{CO}$, needed to produce the measured size of the \co emitting region. If accretion has been steady for longer than one freeze-out time, $L_\mathrm{CO}$ should not be very different from the measured luminosity; on the other hand, if the source has recently undergone an accretion burst, $L_\mathrm{CO}$ will be enhanced relative to the measured luminosity. The ratio, $L_\mathrm{CO}/L_\mathrm{cur}$, can therefore be used to determine if a source has undergone an accretion burst in the past, and to estimate the magnitude of the burst. The calculated $L_\mathrm{CO}/L_\mathrm{cur}$ ratios are listed in Table~\ref{tbl:results} for adopted sublimation temperatures of both \SI{21}{K} and \SI{28}{K}.

\begin{figure}
\includegraphics[width=\hsize]{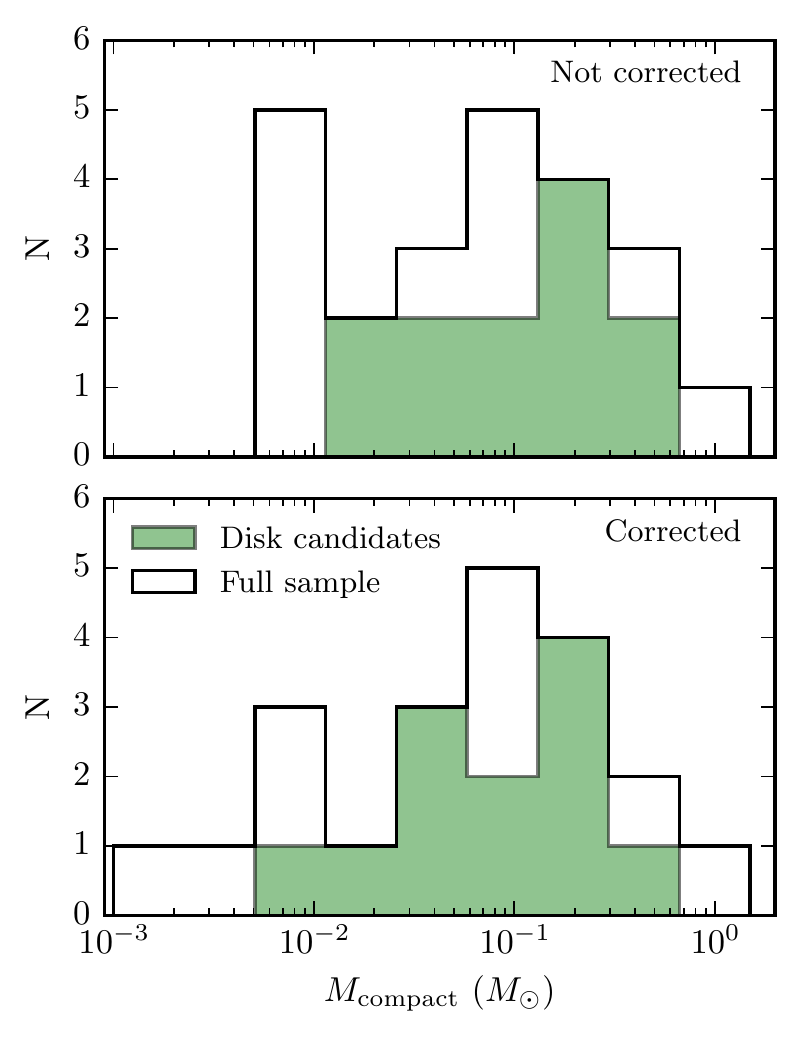}
   \caption{Histograms of compact masses inferred from continuum flux densities at baselines >40\,k$\uplambda$. The black histograms show the full sample while the shaded histograms show only the sources that are identified as disk candidates in Table~\ref{tbl:continuum_results}. The top panel shows uncorrected compact masses, while the bottom panel shows compact masses corrected for the envelope contribution using the method of \citet{Jorgensen:2009bx}.}
   \label{fig:masshist}
\end{figure}

\begin{table*}
\caption{Continuum fluxes and inferred masses.}
\label{tbl:continuum_results}
\centering
\sisetup{separate-uncertainty = true, round-mode= places}
\begin{tabular}{@{\extracolsep{2pt}} l S[table-format=4] S[table-format=4] S[table-format=2.2] S[table-format=1.2] S[table-format=1.2] S[table-format=1.2] S[table-format=1.2] S[table-format=1.2] c l @{}}
\hline\hline
Source & \multicolumn{2}{c}{$F_{1.3\,\mathrm{mm}}$} & \multicolumn{2}{c}{$M_\mathrm{compact}$} & \multicolumn{2}{c}{$S_{850\,\mu\mathrm{m}}$}  & \multicolumn{2}{c}{$M_\mathrm{env}$} & Disk Candidate\tablefootmark{a} & Ref.\tablefootmark{a} \\
\cline{2-3} \cline{4-5} \cline{6-7} \cline{8-9}
 & \multicolumn{2}{c}{(mJy)} & \multicolumn{2}{c}{($M_\sun$)} & \multicolumn{2}{c}{(\si{Jy.beam^{-1}})}  & \multicolumn{2}{c}{($M_\sun$)} & & \\
           &  {u} &  {c} & {u}  &  {c}  &  {u}  & {c}  & {u}  & {c}  &   &      \\
\hline
Per-emb 1  &  115 &   76 & 0.08 &  0.05 &  2.18 & 1.96 & 1.94 & 1.70 & Y & 1,2,6 \\
Per-emb 2  &  415 &  387 & 0.30 &  0.28 &  2.50 & 1.38 & 2.93 & 1.43 & Y & 1,4,5 \\
Per-emb 5  &  207 &  192 & 0.15 &  0.14 &  1.30 & 0.74 & 1.17 & 0.60 & Y & 1,5   \\
Per-emb 11 &  203 &  179 & 0.15 &  0.13 &  1.72 & 1.20 & 1.55 & 1.01 & Y & 1,2   \\
Per-emb 12 & 1690 & 1551 & 1.23 &  1.12 & 11.45 & 6.96 & 8.69 & 4.78 & Y & 7     \\
Per-emb 13 &  725 &  654 & 0.53 &  0.47 &  5.46 & 3.57 & 4.37 & 2.62 &   &       \\
Per-emb 14 &   79 &   59 & 0.06 &  0.04 &  1.18 & 1.01 & 1.30 & 1.08 & Y & 1,2,4 \\
Per-emb 16 &   12 &   -4 & 0.01 & -0.00 &  0.78 & 0.79 & 0.97 & 0.99 &   &       \\
Per-emb 18 &  117 &   85 & 0.08 &  0.06 &  1.85 & 1.60 & 1.24 & 1.04 & Y & 1,5   \\
Per-emb 19 &   10 &    1 & 0.01 &  0.00 &  0.46 & 0.46 & 0.53 & 0.53 &   &       \\
Per-emb 21 &   44 &    8 & 0.03 &  0.01 &  1.85 & 1.83 & 1.24 & 1.22 &   &       \\
Per-emb 22 &   75 &   50 & 0.05 &  0.04 &  1.41 & 1.27 & 0.89 & 0.78 & Y & 4     \\
Per-emb 25 &   83 &   80 & 0.06 &  0.06 &  0.38 & 0.15 & 0.30 & 0.10 &   &       \\
Per-emb 26 &  169 &  132 & 0.12 &  0.10 &  2.23 & 1.85 & 1.11 & 0.88 &   &       \\
Per-emb 27 &  245 &  185 & 0.18 &  0.13 &  3.53 & 2.99 & 1.48 & 1.21 & Y & 1,3   \\
Per-emb 28 &    8 &   -7 & 0.01 & -0.00 &  0.71 & 0.73 & 0.71 & 0.73 &   &       \\
Per-emb 29 &  132 &   85 & 0.10 &  0.06 &  2.60 & 2.35 & 1.84 & 1.64 &   &       \\
Per-emb 33 &  513 &  429 & 0.37 &  0.31 &  5.46 & 4.22 & 3.36 & 2.47 & Y & 1,4   \\
Per-emb 35 &   36 &   19 & 0.03 &  0.01 &  0.86 & 0.80 & 0.35 & 0.33 & Y & 1     \\
Per-emb 42 & {...}& {...}& {...}& {...} &  2.23 & {...}& 1.11 & {...}&   &       \\
Per-emb 44 &  305 &  220 & 0.22 &  0.16 &  4.87 & 4.23 & 1.79 & 1.51 & Y & 1     \\
Per-emb 47 &   10 & {...}& 0.01 & {...} & {...} & {...}& {...}& {...}&   &       \\
Per-emb 53 &   25 &   13 & 0.02 &  0.01 &  0.61 & 0.57 & 0.30 & 0.27 & Y & 1     \\
Per-emb 61 &   10 &    6 & 0.01 &  0.00 &  0.21 & 0.19 & 0.24 & 0.22 &   &       \\
\hline
\end{tabular}
\tablefoot{Columns labelled `c' have been corrected using the method of \citet{Jorgensen:2009bx}, while columns labelled `u' contain the uncorrected values. Typical statistical uncertainties of the compact and extended flux densities, $F_{1.3\,\mathrm{mm}}$ and $S_{850\,\mu\mathrm{m}}$, are \SI{5}{mJy} and \SI{0.05}{Jy.beam^{-1}} respectively. Absolute uncertainties are expected to be around \SI{20}{\percent}. Typical statistical uncertainties of the inferred masses, $M_\mathrm{compact}$ and $M_\mathrm{env}$, are 0.01\,$M_\sun$ and 0.1\,$M_\sun$ respectively. Absolute uncertainties are expected to be on the order of magnitude scale.
\tablefoottext{a}{Disk candidate from resolved continuum observations. References: (1) VANDAM survey, J.\ Tobin, private communication; (2) \citet{SeguraCox:2016kl}; (3) \citet{Tobin:2015gg}; (4) \citet{Tobin:2015fk}; (5) \citet{Tobin:2016fl}; (6) \citet{Lee:2009gp}; (7) \citet{Cox:2015fd}.}
}
\end{table*}

\subsection{Continuum observations}
\label{sec:continuum}

Given the resolution of the data, it is not possible to unambiguously detect circumstellar disks. Instead, continuum observations can be used to measure excess compact emission towards the individual sources, which might then indicate the presence of a disk \citep{Looney:2000hm,Harvey:2003ce,Jorgensen:2005df,Jorgensen:2009bx,Enoch:2011hu}. Here, we use the SMA continuum observations to calculate the compact masses towards the sources in the sample. Single-dish JCMT continuum observations from the SCUBA Legacy Catalogue \citep{DiFrancesco:2008di} are used to calculate the envelope masses and to correct the compact masses for contamination from the envelope.

The compact fluxes are measured by fitting unresolved point sources to the \SI{1.3}{mm} SMA continuum data at baseline s >40\,k$\uplambda$ using Miriad's \texttt{uvfit} routine. At such long wavelengths, the dust emission can be assumed to be optically thin meaning that, for a constant temperature, the dust emission will be proportional to the dust mass. To calculate the total dust plus gas mass of the compact emission, we use the formula
\begin{equation}
  M_\mathrm{compact} = \mathcal{R}_\mathrm{gd} \frac{d^2 F_\nu}{\kappa_\nu B_\nu\left(T_\mathrm{dust}\right)}, \nonumber
\end{equation}
where $F_\nu$ is the measured flux density of the source; $\mathcal{R}_\mathrm{gd}$ is the gas-to-dust mass ratio, assumed to be \num{100}; $d$ is the distance to the source; $T_\mathrm{dust}$ is the dust temperature, assumed to be \SI{30}{K} \citep{Dunham:2014dn}; and the dust opacity, $\kappa_\nu$, is taken to be \SI{0.899}{cm^2.g^{-1}} (\citealt{Ossenkopf:1994tq}; coagulated dust grains with thin ice-mantles at a density of $n_\mathrm{H2}$\,$\sim$\,\SI{e6}{cm^{-3}}, commonly referred to as OH5). Such a mass estimate is associated with a number of uncertainties arising from the assumed gas-to-dust mass ratio, the distance uncertainty, the adopted dust opacity, the assumed dust temperatures, and the degree of contamination from the large-scale envelope on the compact emission, amounting to a total uncertainty approaching an order of magnitude \citep{Dunham:2014dn}.

The sources are embedded within massive envelopes that can be traced with single-dish continuum observations. As with the compact emission, the large-scale emission can be converted into a mass estimate of the envelope. However, on such large scales the physical structure of the envelope, as well as the heating from the central protostar, has to be taken into account. Based on radiative transfer models with power law envelopes, $\rho \propto r^{-1.5}$, \citet{Jorgensen:2009bx} derived an empirical relationship between the source flux and the envelope mass
\begin{equation}
  M_\mathrm{env} = 0.44 \, M_\sun \, \left(\frac{L_\mathrm{bol}}{1\,L_\sun}\right)^{-0.36} \left(\frac{S_{850\,\upmu\mathrm{m}}}{\SI{1}{Jy.beam^{-1}}}\right)^{1.2} \left(\frac{d}{\SI{125}{pc}} \right)^{1.2}, \nonumber
\end{equation}
which is adopted to calculate the envelope mass. Like the compact mass, this quantity is associated with order of magnitude uncertainties arising from the gas-to-dust ratio, the adopted dust opacity, and the assumed envelope structure.

\begin{figure*}
\sidecaption
  \includegraphics[width=12cm]{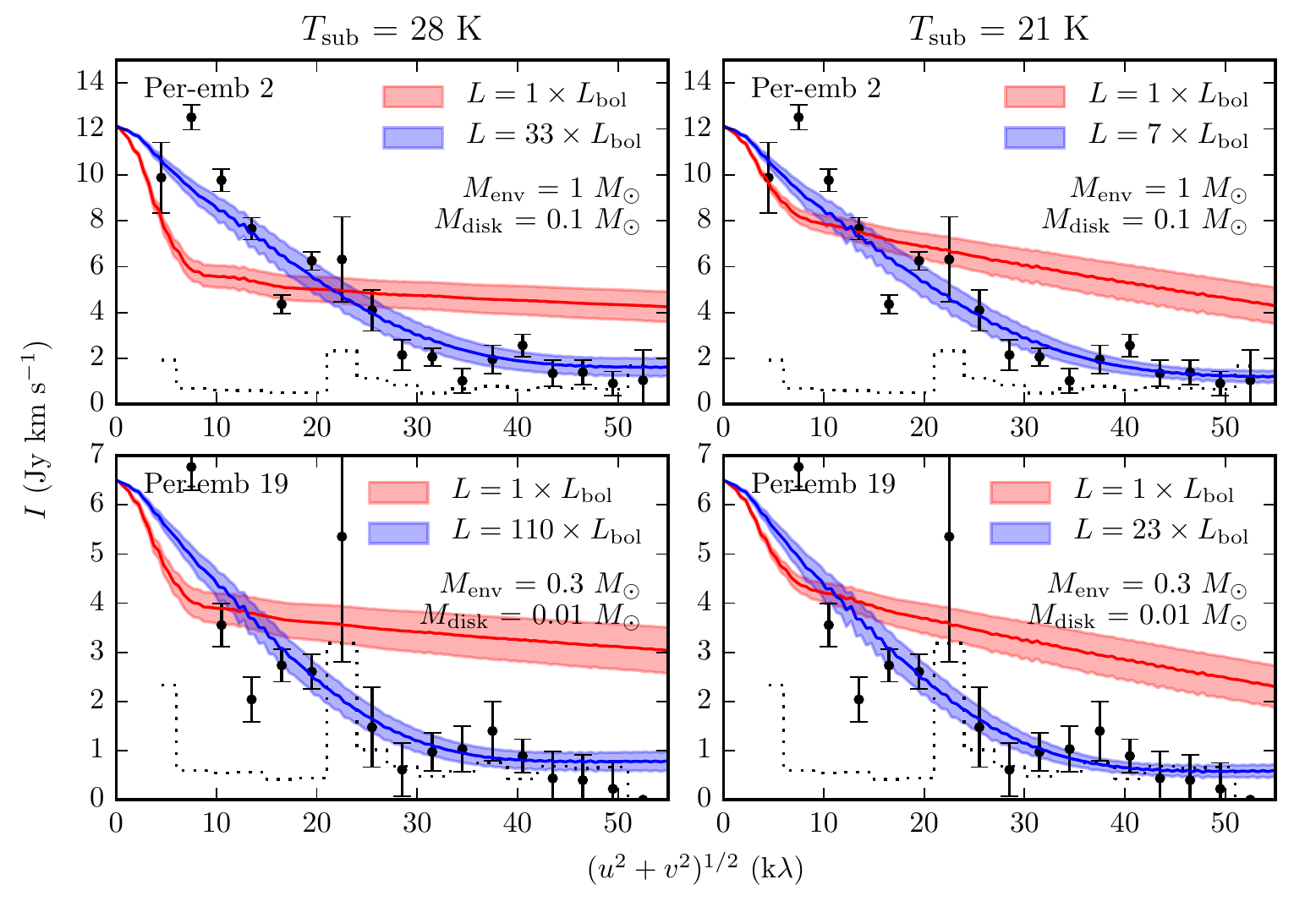}
     \caption{Comparison between observed and calculated $(u,v)$-amplitudes towards Per-emb~2 and~19. The calculated $(u,v)$-amplitudes are obtained from the axissymmetric models presented in Appendix~\ref{sec:appB}. In both cases we adopt an envelope power-law index of 1.5 and outflow half opening angle of \ang{10}. The adopted disk and envelope masses are printed in the panels and were motivated by the measured masses listed in Table~\ref{tbl:continuum_results}. The model visibilities are averaged over 11 viewing angles and are shown as solid lines with shaded 1$\upsigma$ uncertainty regions. To account for the degeneracy between the assumed model masses and CO abundances the model visibilities are normalised to the integrated single-dish flux density of each of the sources. The flux densities were estimated by convolving the moment zero maps with a \ang{;;20} Gaussian beam before reading the flux density at the source position. The red and blue model visibilities show the expected $(u,v)$-amplitudes for the current luminosity and for the luminosity needed to provide a good fit to the measured data. Panels to the left adopt a sublimation temperature of \SI{28}{K} while panels to the right assume a sublimation temperature of \SI{21}{K}.}
     \label{fig:uvamp_modelcomp}
\end{figure*}

The emission from the large-scale envelope is mostly resolved out at baselines >40\,k$\uplambda$, however, depending on the strength of the compact emission, the envelope flux may still contribute significantly (e.g.\ \citealt{Dunham:2014dn}). \citet{Jorgensen:2009bx} calculated that, for a power law envelope with $\rho \propto r^{-1.5}$, approximately \SI{4}{\percent} of the \SI{850}{\micro\metre} single-dish flux leaks into the interferometric flux, measured at \SI{1.1}{mm}, and they therefore subtracted that fraction from their interferometric fluxes. The same method is employed here, except that only \SI{2}{\percent} of the single-dish flux is subtracted, to reflect the fact that the interferometric observations in this study are taken at a wavelength of \SI{1.3}{mm} instead of \SI{1.1}{mm}.

Fluxes and inferred masses are listed in Table~\ref{tbl:continuum_results} along with a column showing whether a given source has been identified as a disk candidate in high-resolution continuum observations. The corrected compact measurements of Per-emb~16 and~28 are negative, meaning that the contamination from the shared envelope between the two sources has been over estimated, which indicates that the real envelope emission is less centrally condensed than assumed (i.e.\ has a shallower power law slope than 1.5). Figure~\ref{fig:masshist} shows the distribution of compact masses. The medians of the corrected and non-corrected mass distributions are 0.06\,$M_\sun$ and 0.08\,$M_\sun$ respectively, and there is no significant difference between the median mass of the full sample and that of the disk candidates. The widths of both distributions span more than an order of magnitude, in agreement with the results of other continuum surveys \citep{Jorgensen:2009bx,Enoch:2011hu,Tobin:2015fk}, and there is no discernible difference between the distributions covering the full sample and the one covering only the disk candidates (a Kolmogorov–Smirnov test yields a \emph{p}-value of \num{0.55}, meaning that the null hypothesis that the two samples are drawn from the same distribution cannot be rejected). This finding indicates that one should be careful about interpreting a high compact mass as evidence of a circumstellar disk, although we cannot rule out that the compact masses are correlated with the disk masses.

\section{Discussion}
\label{sec:discussion}

\subsection{Reliability of measured extents}
\label{sec:reliability}

Given the limited baseline coverage and sensitivity of the observations there is also a limitation on how small \co extents can be measured from the observations. This issue is addressed in Appendix~\ref{sec:appB} where we find that it should be possible to measure the size of the \co emitting region down to angular scales of $\approx$\ang{;;1.4}. In this section we discuss the reliability of the measured extents by comparing the measurements directly to theoretical $(u,v)$-amplitudes as well as to results from other observational studies.

Figure~\ref{fig:uvamp_modelcomp} compares the measured $(u,v)$-amplitudes towards Per-emb~2 and~19 to the calculated $(u,v)$-amplitudes based on the axissymmetric radiative transfer models presented in Appendix~\ref{sec:appB}. These two sources were chosen because both have a low current luminosity (0.9\,$L_\sun$ and 0.4\,$L_\sun$ respectively) and because both show evidence of a past accretion burst. Calculating the $(u,v)$-amplitudes for a model based on the measured bolometric luminosities of the sources yield a poor fit to the data regardless of whether one adopts a CO sublimation temperature of \SI{28}{K} or \SI{21}{K} (red lines in Fig.~\ref{fig:uvamp_modelcomp}). To provide a good fit to the data, the protostellar luminosity of the model has to be increased by the factor indicated by the $L_\mathrm{CO}/L_\mathrm{cur}$ ratios listed in Table~\ref{tbl:results} (blue lines in Fig.~\ref{fig:uvamp_modelcomp}). Based on the comparison between the calculated and observed $(u,v)$-amplitudes we conclude that the large sizes of the C$^{18}$O emitting regions measured towards many of the low-luminosity sources in the sample are real and not a measurement bias of the interferometer.

For bolometric luminosities $\leq$\,3\,$L_\sun$ the predicted size of the \co emitting region is smaller than the smallest measurable size of the \co emitting region, which may explain why all of the low-luminosity sources lie well above the solid blue line in Fig.~\ref{fig:extent}. In Sect.~\ref{sec:c18oobs} four sources, all with $L_\mathrm{bol} \lesssim 1.2\,L_\sun$, were excluded from the sample because the sensitivity of the observations were not high enough to measure the extents reliably. Since the elevation of CO into the gas-phase over a large region is itself expected to increase the strength of the emission, which in turn will make the extent easier to measure, it seems likely that the four excluded sources have not recently undergone an accretion burst and we therefore count them to the `no-burst' group when addressing the burst statistics in Sect.~\ref{sec:lcolcur}.

To test whether the results are consistent over different data sets, the MASSES and \jor measurements are compared to each other. The original \jor extents reported by \citet{Jorgensen:2015kz} were measured using Miriad's \texttt{uvfit} routine. To make a direct comparison possible, the \jor data have been reanalysed using the procedure described in Sect.~\ref{sec:fitting}. A description of the reanalysis along with a comparison to the original results is presented in Appendix~\ref{sec:appA}. The rightmost column of Table~\ref{tbl:results} lists the (re)-measured extents of the \jor sources. The only source whose measured extent is not consistent with its MASSES counterpart within 2$\upsigma$ is Per-emb~13, which has a measured extent of \ang{;;3.9} in the MASSES sample and \ang{;;0.5} in the \jor sample. A closer examination of the fitting results and $(u,v)$-amplitudes (see Fig.~\ref{fig:amplitude_jeskj}) reveals that the \jor emission is consistent with a point source, and that all extents $\lesssim$\ang{;;2} fit the data well. While closer, it is still significantly less than the measured extent of the MASSES source.

The sizes of the \co emitting regions towards Per-emb~12, 13, and 26 have also recently been measured by \citet{Anderl:2016iv} using sub-arcsecond observations taken with the IRAM Plateau de Bure Interferometer. Their measurements, obtained by fitting two-dimensional Gaussians to maps of integrated emission, are in excellent agreement with our own and are given as: \SI[separate-uncertainty]{5.4(1)}{arcsec} for Per-emb~12; \SI[separate-uncertainty]{3.3(1)}{arcsec} for Per-emb~13; and \SI[separate-uncertainty]{3.5(2)}{arcsec} for Per-emb~26.

\subsection{Emission asymmetry}
\label{sec:asymmetry}

On average, the ratio between the major and minor axes of the Gaussian fits listed in Table~\ref{tbl:results} is 1.5 with individual values ranging from 1.06 to 2.5 (accepted fits only). The emission may appear elongated for several reasons; the gas and dust surrounding the individual sources likely deviate from spherical symmetry, which can influence the morphology of the emission; some sources (e.g. Per-emb~35 and~44) contain several protostars that are close enough together that they are not resolved by the interferometer, but far enough apart that the shape of the emission is affected; finally, the emission may be elongated in the direction of the outflow. In the first two cases it is appropriate to use the method applied in this study and calculate an average size of the \co emitting region from the minor and major axes of the fitted Gaussian. However, if the emission is elongated along the outflow axis, it may be more appropriate to take the FWHM in the direction perpendicular to the outflow axis because the temperatures in the outflow cone are significantly enhanced and CO therefore sublimates at larger radii.

\begin{table}
\caption{Outflow and Gaussian position angles.}
\label{tbl:outflow}
\centering
\begin{tabular}{l S S S l}
\hline\hline
Source & {Outflow P.A.} & {Gauss P.A.} & {Rel. P.A.} & Ref. \\
       & {(degree)}   & {(degree)} & {(degree)} & \\
\hline
Per-emb  1    & 116 & 126 & 10 & 1,2 \\
Per-emb  2    & 129 & 150 & 21 & 2   \\
Per-emb  5    & 125 & 100 & 25 & 2   \\
Per-emb 11,O1 & 161 &   2 & 21 & 3   \\
Per-emb 11,O2 &  36 &   2 & 34 & 3   \\
Per-emb 12    &  35 & 145 & 70 & 4   \\
Per-emb 13    & 176 & 179 &  3 & 3   \\
Per-emb 14    &  95 &  78 & 17 & 5   \\
Per-emb 16    &   7 &  26 & 19 & 3   \\
Per-emb 18    & 150 & 127 & 23 & 3   \\
Per-emb 19    & 148 &  64 & 84 & 2   \\
Per-emb 21    &  48 &  87 & 39 & 3   \\
Per-emb 22    & 118 &  46 & 72 & 2   \\
Per-emb 26    & 162 & 145 & 17 & 3   \\
Per-emb 27,O1 &  14 &  24 & 10 & 4   \\
Per-emb 27,O2 & 104 &  24 & 80 & 4   \\
Per-emb 29    & 132 & 178 & 46 & 2   \\
Per-emb 33    & 122 &  22 & 80 & 3   \\
Per-emb 35,O1 & 123 &  52 & 71 & 2   \\
Per-emb 35,O2 & 169 &  52 & 63 & 2   \\
Per-emb 44    & 130 &  11 & 61 & 4   \\
Per-emb 53    &  59 &  15 & 44 & 2   \\
\hline
\end{tabular}
\tablefoot{The first two columns report the outflow and Gaussian position angles measured from North to East. The third column gives the difference between the two position angles and is calculated such that the reported angle is always $\leq$\ang{90}. Per-emb~11, 27, and 35 drive two outflows for which both position angles are listed. References: (1) \citet{McCaughrean:1994jm}; (2) Stephens et al. (in preparation); (3) \citet{Lee:2016fh}; (4) \citet{Plunkett:2013cc} (5) \citet{Tobin:2015fk}.}
\end{table}

\begin{figure}
\includegraphics[width=\hsize]{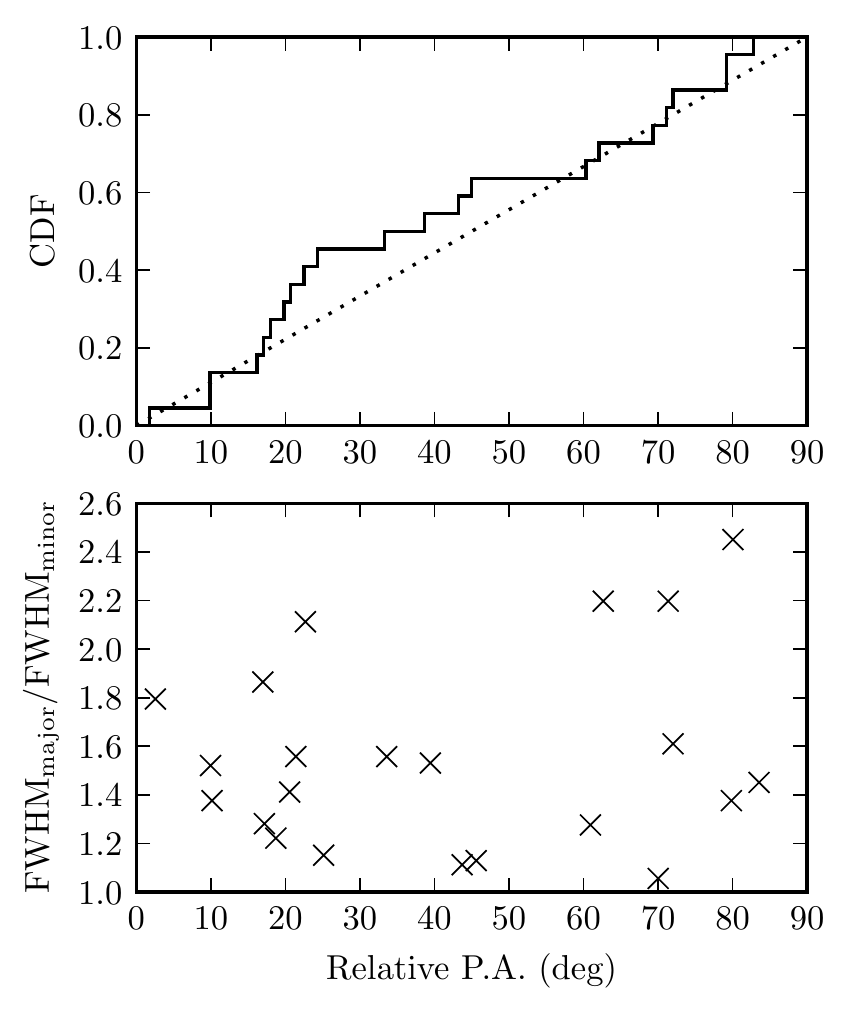}
   \caption{Top: Cumulative distribution function of the difference between the Gaussian and outflow position angles. The dotted line shows a uniform distribution expected for a random relative orientation. Bottom: $\mathrm{FWHM}_\mathrm{major}/\mathrm{FWHM}_\mathrm{minor}$ vs. relative position angle.}
   \label{fig:relative_angle}
\end{figure}

We investigate whether the position angles of the fitted Gaussians tend to be aligned with the outflow axes. Table~\ref{tbl:outflow} lists outflow and Gaussian position angles for all sources in the sample where the Gaussian fit was accepted and where we could find a measurement of the outflow position angle in the literature. The top panel of Fig.~\ref{fig:relative_angle} shows the cumulative distribution of the relative position angles between Gaussians and outflows (solid line) along with the uniform cumulative distribution (dotted line). Performing a Kolmogorov-Smirnov test on the distribution with null hypothesis that the observed and random distributions are the same yields a \emph{p}-value of 0.46. Adopting the canonical significance threshold of 0.05, the null hypothesis can therefore not be rejected, and the measurements are consistent with being drawn from a distribution with no preferred alignment. In the bottom panel of Fig.~\ref{fig:relative_angle} we plot the ratio between the Gaussian major and minor axes against the relative position angle. The aim of this is to investigate whether more elongated emission has a stronger tendency towards alignment with the outflow axis. The figure reveals no such correlation and we therefore conclude that there is no general tendency towards alignment between the outflow axes and the Gaussian major axes.

While we reject a general connection between elongated emission and outflow axes, there are a few individual sources where it appears likely that the elongation is caused by the outflow. This is the case for Per-emb~1, 13, and~27, all of which are known to drive strong outflows and all of which have relative position angles $\leq$\,\ang{10}. Among the three sources $FWHM_\mathrm{avg}$ and $FWHM_\mathrm{minor}$ differ by \SI{34}{\percent} in the most extreme case. Changing the reported extent from $FWHM_\mathrm{avg}$ to $FWHM_\mathrm{minor}$ may thus affect the conclusions for the individual sources but will not affect the conclusions of the sample as a whole.

\citet{Frimann:2016ke} studied synthetic C$^{18}$O maps towards a sample of 2000 embedded protostars in a large three-dimensional MHD simulation of a molecular cloud, and fitted Gaussians to their synthetic data in the same manner as is done in this study. On average, they found a ratio between the major and minor axes of the fitted Gaussians of 1.2; somewhat below the average value of 1.5 found in this study. However, the authors did not add noise to their synthetic observables nor did they adopt a realistic sampling of the $(u,v)$-plane, which may explain the difference. We have therefore taken the synthetic data from the numerical study and applied noise and $(u,v)$-sampling similar to what is available in the observations (see Sect.~\ref{sec:Bco} for details). Calculating the average ratio between the Gaussian major and minor axes of the new data gives a value of 1.6, which is similar to the 1.5 measured here. It is worth noting that while the resolution of the MHD simulation studied by \citet{Frimann:2016ke} was large enough to resolve the protostellar cores, it was not large enough to resolve the launching region of protostellar outflows. The study also did not include any multiple systems, which means that the elongation measured in the Gaussian fits must be due to deviations from spherical symmetry of the material surrounding the protostars, as well as overall uncertainties introduced when adding noise and a realistic $(u,v)$-sampling to the synthetic observations.

\subsection{$L_\mathrm{CO}/L_\mathrm{cur}$ as a tracer of past accretion bursts}
\label{sec:lcolcur}

The main tracer used for detecting past accretion bursts is the ratio, $L_\mathrm{CO}/L_\mathrm{cur}$, the calculation of which is associated with a number of uncertainties. These arise from statistical and systematic measurement uncertainties of the \co extents and bolometric luminosities, as well as from uncertainties and assumptions pertaining to the models that went into calculating Eq.~\eqref{eq:extent}. While statistical uncertainties are accounted for when calculating $L_\mathrm{CO}/L_\mathrm{cur}$, systematic and model uncertainties are not. Some of these uncertainties are discussed in Appendix~\ref{sec:appB}, which presents the radiative transfer models that were used to derive Eq.~\eqref{eq:extent}. The appendix examines how the physical structure of the disk, envelope and outflow affects the measured luminosities and \co extents. Other uncertainties, such as $L_\mathrm{bol}$ uncertainties originating from incomplete sampling of the SED or calibration uncertainties of the observations, are more difficult to track, but the fact that the studied sample consists of more than 20 sources helps even those out.

A key assumption concerning Eq.~\eqref{eq:extent} regards the temperature at which CO ice sublimates from the dust grains. In the calculation leading up to Eq.~\eqref{eq:extent} two sublimation temperatures were considered: $T_\mathrm{sub} = \SI{28}{K}$ appropriate for CO ice mixed with water \citep{Noble:2012hq}, and $T_\mathrm{sub} = \SI{21}{K}$ appropriate for pure CO ice \citep{Sandford:1993da,Bisschop:2006dc}. A low sublimation temperature would explain the extended \co emission measured towards some sources, but it struggles to explain the compact emission observed towards others (cf. Fig.~\ref{fig:extent}). Observationally, there is some evidence that the sublimation temperature of CO in cores is higher than the \SI{21}{K} expected for pure CO ice, with studies reporting values in the range \SIrange[range-phrase=--]{24}{40}{K} \citep{Jorgensen:2005cs,Yldz:2012cz,Yldz:2013bt,Anderl:2016iv}. Studies of disks have found even lower CO sublimation temperatures of \SIrange[range-phrase=--]{17}{19}{K} \citep{Qi:2013dx,Mathews:2013jb}. However, the different sublimation temperatures inferred for the disks may be explained by thermal processing of the disk material, which can create an `onion shell' structure of the ices on the dust grains with separated layers of pure ices \citep{Jorgensen:2015kz}.

In general, one can consider the evidence of a past accretion burst to grow stronger as the ratio $L_\mathrm{CO}/L_\mathrm{cur}$ increases. If we follow \citet{Jorgensen:2015kz} and only consider sources with $\left(L_\mathrm{CO}/L_\mathrm{cur}\right)_{28\,\mathrm{K}} > 5$, 12 out of 23 sources (\SI{52}{\percent}) show evidence of a past accretion burst. This statistic excludes Per-emb~42, whose emission could not be disentangled from that of Per-emb~26, but includes Per-emb~25, 28, 47, and 61 which we consider no-burst sources (cf. Sect.~\ref{sec:reliability}). Considering $\left(L_\mathrm{CO}/L_\mathrm{cur}\right)_{21\,\mathrm{K}} > 5$ the statistics change so that four out of 23 sources (Per-emb~2, 11, 16, and 19; \SI{17}{\percent}) continue to show evidence of a past accretion burst. In either case, a significant fraction of the sources show evidence of having undergone an accretion burst in the not so distance past. We note that even though the $L_\mathrm{CO}/L_\mathrm{cur}$ ratio gives an indication of the burst magnitude it is not constant with time but will start decreasing as CO begins to refreeze. Also, the exclusion of baselines below 15\,k$\uplambda$ means that we are unable to measure large extents, which may lower the measured $L_\mathrm{CO}/L_\mathrm{cur}$ ratios towards some of the high-luminosity sources. Because of these uncertainties, we consider all sources with $L_\mathrm{CO}/L_\mathrm{cur} > 5$ to show evidence of a past accretion burst even though classical FUor accretion bursts show luminosity enhancements of an order of magnitude or greater.

\subsection{Burst intervals}
\label{sec:burstinterval}

The time it takes for the molecules to freeze back onto the dust grains following a burst is inversely proportional to the density \citep{Rodgers:2003dg}. The length of the time interval where it is still possible to observe the spatially extended C$^{18}$O emission therefore depends both on the magnitude of the burst (high-intensity bursts push the CO ice-line out to larger radii where the freeze-out time scale is longer) and on the physical structure of the envelope. Knowledge of the freeze-out time scale, $t_\mathrm{dep}$, can be used to estimate the time interval between bursts, $T_\mathrm{burst}$. Assuming that the duration of the burst itself is negligible, the average interval between bursts can be estimated by noting that the ratio $t_\mathrm{dep}/T_\mathrm{burst}$ should be equal to the fraction of sources that show evidence of having undergone an accretion burst in the past. For a sublimation temperature of \SI{28}{K} this suggests $T_\mathrm{burst} \approx 2 \times t_\mathrm{dep}$ and for a sublimation temperature of \SI{21}{K} it suggests $T_\mathrm{burst} \approx 5 \times t_\mathrm{dep}$. For an average freeze-out time scale of \SI{10000}{yr}, appropriate for an envelope density of \SI{e6}{cm^{-3}} \citep{Rodgers:2003dg,Visser:2012dp}, a rough estimate of the burst interval is thus \SIrange[range-phrase=--]{20000}{50000}{yr}.

\citet{Scholz:2013ij} used a direct observational approach to estimate the average burst interval of YSOs. Utilising mid-infrared Spitzer and WISE photometry, they identified five burst candidates in a sample of \num{4000} YSOs observed over two epochs set five years apart. Based on their analysis, \citet{Scholz:2013ij} argued for an average burst interval in the range from \SIrange{5000}{50000}{yr}. A different approach was adopted by \citet{Offner:2011ex} who used the local star formation rate and number of identified FUors to infer that YSOs spend $\sim$\SI{1200}{yr} undergoing accretion bursts. Assuming a \SI{100}{yr} burst life-time and an average protostellar lifetime of \SI{0.5}{Myr}, the time between bursts is $\sim$\SI{40000}{yr}. Both these observational estimates are in good agreement with the rough estimate from above. Estimates of burst intervals can also be gained from numerical simulations. Using hydrodynamical simulations, \citet{Vorobyov:2015dv} found that episodic accretion events induced by gravitational instabilities and disk fragmentation are present during the early evolution of most protostellar systems. The average burst interval range from \SIrange{0.3e4}{1.1e4}{yr} depending on the initial conditions of the model; somewhat below but at the same scale as the estimate above.

To accurately estimate the average burst interval, observations over long time baselines and/or large sample sizes are needed. For example, for a hypothetical average burst interval of \SI{20000}{yr} and a sample size consisting of \num{4000} sources, a time baseline $\gtrsim$\SI{30}{yr} is needed to constrain the length of the burst interval to a factor of two or better. Similarly, for a time baseline of \SI{5}{yr} a sample size $\gtrsim$\num{20000} is needed to constrain the length of the burst interval \citep{Hillenbrand:2015bx}. Such large surveys are challenging to perform, particularly for deeply embedded Class~0 and~I protostars that are only detected at far-infrared wavelengths and beyond. One of the advantages of using freeze-out and sublimation chemistry to trace past accretion bursts is that the effective time baseline for the observations is equal to the freeze-out time of the molecules. With freeze-out times generally expected to exceed \SI{e3}{yr}, it should therefore be possible to constrain the burst interval to within a factor of two, with a sample comprised of tens of objects instead of thousands of objects. To actually do this requires careful modelling of the envelope structure to better constrain the freeze-out time scale, as well as observations of a molecule other than CO, with another freeze-out time scale, to further decrease uncertainties. A good candidate molecule for follow-up observations is formaldehyde (H$_2$CO), which has a sublimation temperature, $T_\mathrm{sub} = \SI{38}{K}$ \citep{Aikawa:1997hk}, that is low enough that the extent of its emission should still be measurable with an interferometer such as the SMA.

\subsection{Link to circumstellar disks}
\label{sec:linkdisk}

Given that episodic outbursts are thought to arise from instabilities in circumstellar disks, it is of particular interest to investigate whether evidence of a past accretion burst can be linked to the presence of a circumstellar disk. The majority of the sources in the observed sample drive outflows, which already indicate that they are encircled by a disk, but not necessarily a large or massive one. A classical FUor accretion burst, with an accretion rate of \num{e-4}\,$M_\sun\,\mathrm{yr}^{-1}$ lasting \SI{200}{yr}, will accrete \num{0.02}\,$M_\sun$ onto the protostar, indicating that the disk has to be fairly massive to sustain such a burst. Furthermore, for gravitational instabilities to operate, disk masses $\gtrsim$\num{0.05}\,$M_\sun$ are needed \citep{Vorobyov:2013dl,Kratter:2016tj}. Although magnetic and thermal instabilities may continue to work in the inner disk independently of the total disk mass, gravitational instabilities in the outer disk are needed to feed this region \citep{Armitage:2001jl,Zhu:2009fv}. Consequently, there is some reason to believe that protostars encircled by massive disks are the most likely burst candidates.

The unambiguous detection of a circumstellar disk requires high-resolution line observations for measuring Keplerian rotation; something that remains challenging for Class~0 sources, for which only four detections have been reported so far \citep{Tobin:2012ee,Murillo:2013eq,Codella:2014hy,Lindberg:2014bq}. In place of line observations, disk candidates can be identified from continuum observations, either through resolved observations (e.g.\ \citealt{Tobin:2015fk,SeguraCox:2016kl}) or through the detection of excess compact emission (e.g.\ \citealt{Jorgensen:2009bx}). Relating disk candidates from resolved observations to the compact masses measured in this work has revealed no discernible pattern (Fig.~\ref{fig:masshist}). Specifically, sources with large compact masses do not appear to be more likely to be identified as disk candidates from resolved continuum observations. This indicates that simply detecting an excess of compact emission on small scales is not enough to claim the detection of a disk. The excess emission may then simply originate from an inner over-dense region of the envelope, such as a magnetically supercritical core, which may form during the collapse \citep{Chiang:2008jp}.

\begin{figure}
\includegraphics[width=\hsize]{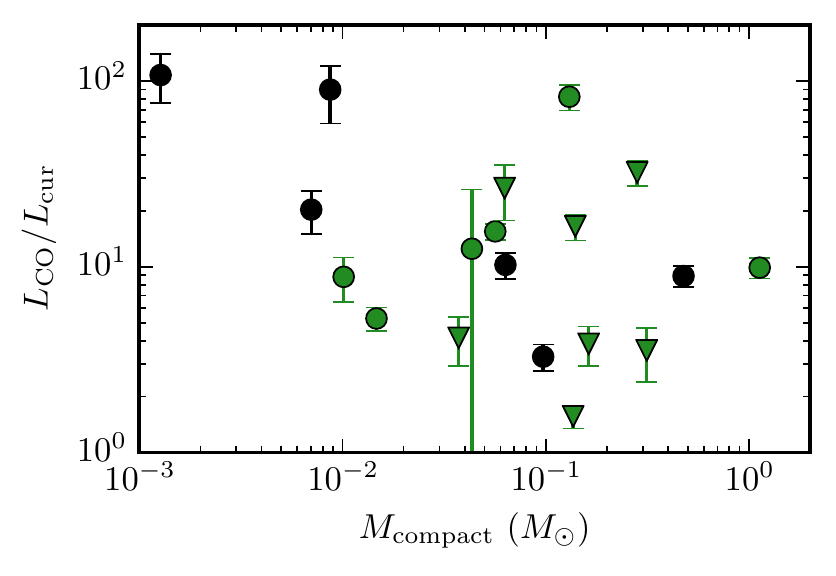}
   \caption{Corrected $M_\mathrm{compact}$ (column marked `c' in Table~\ref{tbl:continuum_results}) vs.\ $\left(L_\mathrm{CO}/L_\mathrm{cur}\right)_{28\,\mathrm{K}}$. The corrected compact mass of Per-emb~16 is negative, and the uncorrected mass has therefore been used for this object. Green symbols represent sources that are identified as disk candidates in resolved continuum surveys (cf.\ Table~\ref{tbl:continuum_results}). Triangular symbols represent sources that are associated with a companion protostar at a projected distance <\ang{;;0.8} (\citealt{Tobin:2016fl}; also see Fig.~\ref{fig:c18omap}). Round symbols represent sources with no close companion.}
   \label{fig:MdiskvsLcoLcur}
\end{figure}

Figure~\ref{fig:MdiskvsLcoLcur} shows compact masses vs.\ $\left(L_\mathrm{CO}/L_\mathrm{cur}\right)_{28\,\mathrm{K}}$ for the objects in the sample. No significant correlation between the compact mass and $\left(L_\mathrm{CO}/L_\mathrm{cur}\right)_{28\,\mathrm{K}}$ is seen in the figure, and several of the sources with small compact masses simultaneously show $\left(L_\mathrm{CO}/L_\mathrm{cur}\right)_{28\,\mathrm{K}}$ ratios significantly larger than \num{10}, in apparent disagreement with the idea that objects encircled by massive disks are the most likely burst candidates. Two of the sources with small compact masses, Per-emb~16 and~19, simultaneously have very large $\left(L_\mathrm{CO}/L_\mathrm{cur}\right)_{28\,\mathrm{K}}$ ratios of $\sim$\num{100}. Using ALMA Cycle~0 observations, \citet{Jorgensen:2013ej} observed a similar situation towards the Lupus~I embedded protostar, IRAS~15398--3359, which also shows evidence of a past high-intensity burst, while lacking strong compact emission. \citeauthor{Jorgensen:2013ej} argued that the disk may have been expended during the burst, which can also explain the lack of compact emission seen towards Per-emb~16 and~19. To expend the entire disk in one burst would require an accretion rate of $\sim$\num{e-4}\,$M_\sun\,\mathrm{yr}^{-1}$, corresponding to a burst luminosity $\gtrsim$\num{e2}\,$L_\sun$. This is somewhat more than the estimated burst luminosities, $L_\mathrm{CO}$, of Per-emb~16 and~19, which are both $\sim$\num{40}\,$L_\sun$, however the estimate of $L_\mathrm{CO}$ is highly uncertain, so the scenario cannot be ruled out.

Excluding Per-emb~16 and~19, only one other source, Per-emb~21, for which we see evidence of a past intense burst ($\left(L_\mathrm{CO}/L_\mathrm{cur}\right)_{28\,\mathrm{K}} > 10$), has $M_\mathrm{compact} < 0.05\,M_\sun$. This could suggest a connection between episodic outbursts and massive disks. However, the number statistics are small and the compact mass measurements are highly uncertain. High-resolution and high-sensitivity observations with ALMA, which will be able to better disentangle the emission from the disk and envelope, are needed to get a better grasp of the link between episodic outbursts and disks.

\subsection{Link to multiple systems}

In addition to bursts induced by disk instabilities, it has also been suggested that bursts can be induced by external interactions with one or more companion stars. The idea was first put forth by \citet{Bonnell:1992fk}, who suggested that FUor outbursts may be due to a perturbation of the circumstellar disks as the two members of a close binary system on an eccentric orbit pass each other at periastron. The idea garnered further support due to the fact that several of the known FUors are known to be binaries, among those FU~Orionis itself \citep{Wang:2004di}. Using FU~Orionis as a case study, \citet{Reipurth:2004jk} proposed a paradigm where FUor outbursts are a consequence of the break-up of a non-hierarchical triple system, where one object is being ejected, while the two remaining objects spiral in towards each other, perturbing their disks in the process.

With the results from the VANDAM survey of \citet{Tobin:2016fl}, we have access to the full binary statistics of the protostars in Perseus down to spatial scales of $\sim$\SI{15}{AU}. Of the 24 MASSES objects studied in this paper, 16 (\SI{67}{\percent}) are part of a multiple system with projected separations ranging from \ang{;;0.08} to \ang{;;16} (\SI{19}{AU} to \SI{3760}{AU}). Of the eight objects in the sample that have $\left(L_\mathrm{CO}/L_\mathrm{cur}\right)_{28\,\mathrm{K}} > 10$, six are part of a multiple system. Of these, Per-emb~2, 5, and~18 are all close binary systems (see Fig.~\ref{fig:MdiskvsLcoLcur}) with projected distances between \ang{;;0.080} and \ang{;;0.097} (\SI{19}{AU} and \SI{23}{AU}), and these are the three closest binary systems identified by VANDAM. While this indicates that binarity is not a prerequisite for episodic accretion, the fact that evidence of episodic accretion is observed towards the three closest binaries in the VANDAM sample also suggests that interactions between close binaries may play an important role for episodic outbursts.

\section{Summary}
\label{sec:Summary}

SMA observations of continuum and C$^{18}$O\,$J$\,=\,2--1 line emission have been presented towards a sample of 24 embedded protostars in the Perseus molecular cloud. The aim of the paper has been to use sublimation and freeze-out chemistry of CO to study how accretion proceeds in the earliest, most deeply embedded stages of star formation. The main findings of the paper are as follows
\begin{enumerate}
  \item We have fitted two-dimensional Gaussians to the integrated \co visibilities to measure the sizes of the \co emitting regions towards the protostars in the sample. We succeed except for one case where the signal cannot be properly untangled from the emission of a nearby luminous protostar, and four cases where the signal is not strong enough to obtain a reliable measurement. In the latter case we argue that the sources are likely to not have undergone an accretion burst in the recent past since a burst would have enhanced the signal, thereby making the measurement possible.
  \item Using radiative transfer models, we calculate the predicted size of the \co emitting region as a function of luminosity for adopted CO sublimation temperatures of \SI{21}{K} and \SI{28}{K} (appropriate for pure and water-mixed ices respectively). For $T_\mathrm{sub} = \SI{28}{K}$ the sizes of the emission regions are measured to be significantly enhanced (by more than a factor of five in the calculated burst luminosity, $L_\mathrm{CO}$, relative to the current luminosity, $L_\mathrm{cur}$) for \SI{50}{\percent} of the sources. For $T_\mathrm{sub} = \SI{21}{K}$ the fraction is smaller, \SI{20}{\percent}, but regardless of the adopted sublimation temperature the results indicate that accretion variability is common, even in the earliest stages of star formation.
  \item Given the fraction of sources that show an extended morphology of their \co emission, we estimate an average burst interval of \SIrange[range-phrase=--]{20000}{50000}{yr} depending on the adopted sublimation temperature. Although it is only a rough estimate, it is in agreement with previous estimates from both observations and theory.
  \item Compact \SI{1.3}{mm} continuum fluxes are measured towards the sampled sources and corrected for the contribution from the surrounding envelope to get an estimate of the disk mass of the individual sources. The compact masses show no correlation with the disk candidates inferred from resolved continuum observations, which indicates that one should be careful with interpreting excess unresolved emission from embedded protostars as an indication of the presence of a disk.
  \item Theoretically, accretion bursts can be linked to the presence of a massive disks around the YSO. We attempt to link the observational evidence of past accretion bursts (from the \co observations) to evidence of massive disks (from the continuum observations), but cannot establish a clear link. While this could indicate that bursts and massive disks are not as intimately connected as suggested by theory, a more likely interpretation probably is that a better tracer of the disk mass than the compact continuum emission is needed.
  \item The three closest binaries in the observed sample (projected separations <\,\SI{20}{AU}) all show evidence of extended C$^{18}$O emission, suggesting a tentative connection between binarity and episodic accretion. Several systems also show extended C$^{18}$O emission without being part of a known close binary, indicating that episodic accretion events are not likely to be driven solely by interactions between companions.
\end{enumerate}
In the presented analysis, we have adopted two values of the CO sublimation temperature corresponding to pure and water-mixed CO ice respectively. In both cases, a significant fraction of the sources show evidence of past accretion variability. However, one important uncertainty to the conclusions regards the correct value of the sublimation temperature and whether it is subject to large source-to-source variations. If source-to-source variations turn out to be important, much of the extended \co emission may be explained, not by a past accretion burst, but by a varying sublimation temperature. To address this question follow-up observations of another molecule, expected to show similar freeze-out and sublimation behaviour relative to CO, must be performed. By analysing the new data, one may then investigate whether the same type of behaviour is observed for the new molecule as it was for CO. A prime candidate is formaldehyde (H$_2$CO), which has a sublimation temperature of \SI{38}{K}, and should therefore be possible to resolve by an interferometer, like the SMA. Regardless of the uncertainties on the sublimation temperature, the fact that a significant fraction of the observed protostars show evidence of past accretion variability for a sublimation temperature as low as \SI{21}{K} is a strong indication that accretion is variable, even in the earliest stages of star formation.

\begin{acknowledgements}

The authors are grateful to the anonymous referee for constructive criticism, which led to a significant improvement of the paper. This work is based primarily on observations obtained with the SMA, a joint project between the Smithsonian Astrophysical Observatory and the Academia Sinica Institute of Astronomy and Astrophysics and funded by the Smithsonian Institution and the Academia Sinica. The authors thank the SMA staff for executing these observations as part of the queue schedule, and Charlie Qi and Mark Gurwell for their technical assistance with the SMA data. This research was supported by a Lundbeck Foundation Group Leader Fellowship as well as the European Research Council (ERC) to JKJ under the European Union's Horizon 2020 research and innovation programme (grant agreement No 646908) through ERC Consolidator Grant ``S4F''. EIV acknowledges support from RFBR grant 17-02-00644. Research at Centre for Star and Planet Formation is funded by the Danish National Research Foundation and the University of Copenhagen's programme of excellence.

\end{acknowledgements}

\appendix

\section{Reanalysis of \jor sample}
\label{sec:appA}

\begin{figure*}
\centering
\includegraphics[width=18cm]{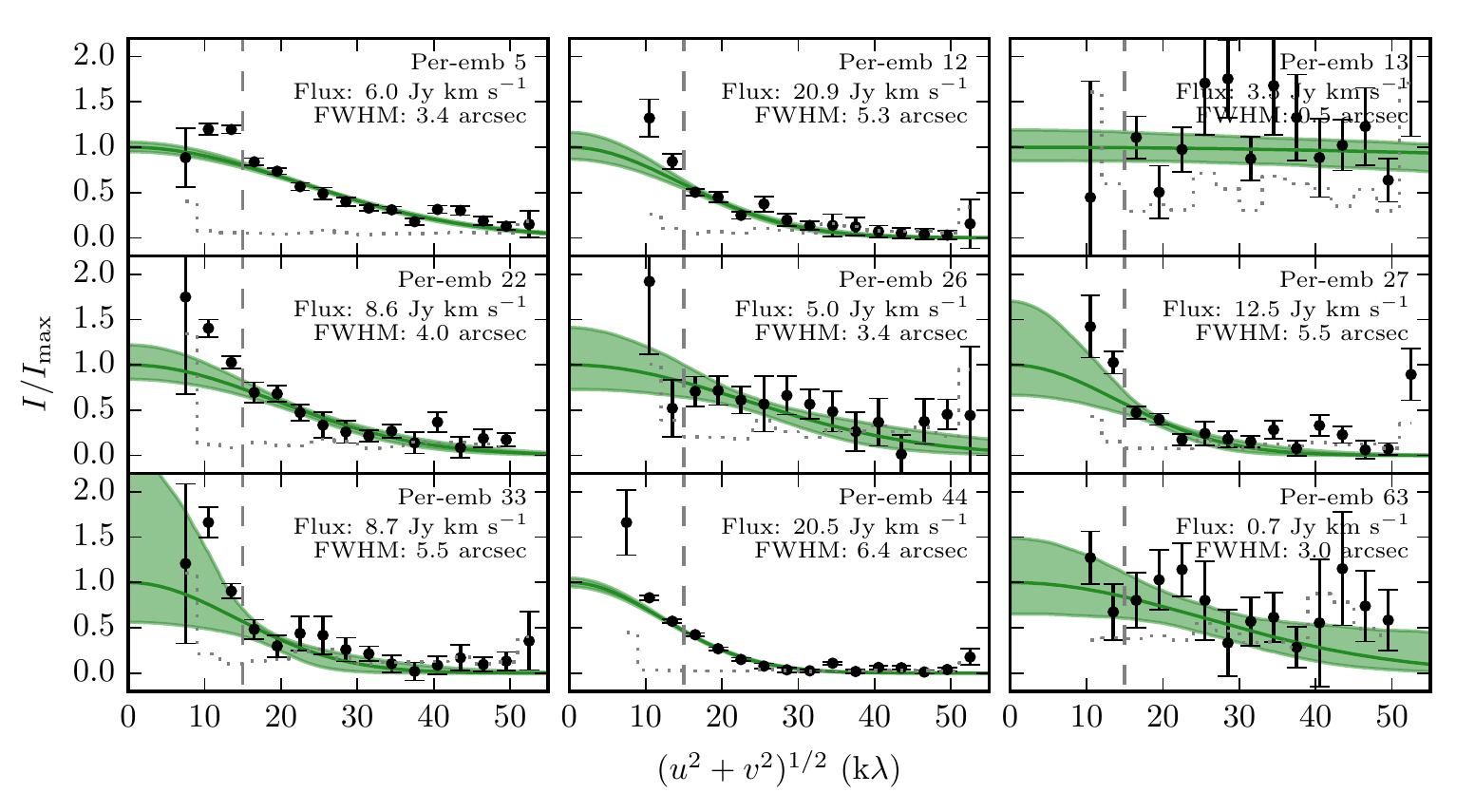}
   \caption{$(u,v)$-amplitude plots of the integrated C$^{18}$O emission towards the sources in the \jor sample. The solid green line in each panel is the best fitting Gaussian, while the shaded region indicate the 1$\upsigma$ uncertainty region of the fit. The vertical axis of each panel has been scaled to the peak flux of the Gaussian fit, whose value is printed in the upper right. The grey dashed lines indicate the 15\,k$\uplambda$ lower boundary of the fit.}
   \label{fig:amplitude_jeskj}
\end{figure*}

This appendix presents the results of the reanalysis of the nine Perseus sources from the \jor sample, eight of which are also in the MASSES sample analysed in the main paper. The sources have been analysed the same way as presented in Sects.~\ref{sec:fitting} and~\ref{sec:results}. Here, we compare our analysis of the \jor sample (Table~\ref{tbl:appA}) to the original results presented in \citet{Jorgensen:2015kz}, which were obtained using Miriad's \texttt{uvfit} routine. $(u,v)$-amplitude plots are shown in Fig.~\ref{fig:amplitude_jeskj}. For a detailed description of the original analysis and of the data reduction, the reader is referred to \citet{Jorgensen:2015kz}.

The original sizes of the \co emitting regions reported by \citet{Jorgensen:2015kz} are overall consistent with the results of the reanalysis. Some deviations are expected since the original analysis assumed circular Gaussians and utilised Miriad's \texttt{uvfit} routine, while the reanalysis assumed two-dimensional Gaussians and utilised an MCMC algorithm. Still, assuming an average statistical uncertainty of \ang{;;0.5} on the measurements reported by \citet{Jorgensen:2015kz}, all are consistent with the results of the reanalysis within 2$\upsigma$.

\begin{table*}
\caption{Gaussian fitting results for \jor sample.}
\label{tbl:appA}
\centering
\begin{tabular}{l S[separate-uncertainty,table-format=2.1(1)] S[table-format=1.1]@{\hskip 2pt}l S[table-format=2.2]@{\hskip 2pt}l | S[table-format=1.1]@{\hskip 2pt}l S[table-format=1.1]}
\hline\hline
Source & {$L_\mathrm{bol}$} & \multicolumn{2}{c}{$\mathrm{FWHM}_\mathrm{avg}$} & \multicolumn{2}{c|}{$\mathrm{FWHM}_\mathrm{major}/\mathrm{FWHM}_\mathrm{minor}$} & \multicolumn{2}{c}{$\mathrm{FWHM}_\mathrm{avg,MASSES}$} & {$\mathrm{FWHM}_\mathrm{avg,Miriad}$} \\
 & {($L_\sun$)} & \multicolumn{2}{c}{(arcsec)} & & & \multicolumn{2}{c}{(arcsec)} & {(arcsec)} \\
\hline
Per-emb 5  & 1.3(1)  &  3.4 & $\substack{+0.1 \\ -0.1}$    & 1.06 & $\substack{+0.06 \\ -0.04}$  &  3.2 & $\substack{+0.2 \\ -0.2}$    & 4.0 \\[2pt]
Per-emb 12 & 7.0(7)  &  5.3 & $\substack{+0.5 \\ -0.4}$    & 1.13 & $\substack{+0.10 \\ -0.08}$  &  5.1 & $\substack{+0.2 \\ -0.2}$    & 5.5 \\[2pt]
Per-emb 13 & 4.0(3)  &  0.5 & $\substack{+0.6 \\ -0.3}$    &   12 & $\substack{+37 \\ -9}$       &  3.9 & $\substack{+0.2 \\ -0.2}$    & 1.9 \\[2pt]
Per-emb 22 & 3.6(5)  &  4.0 & $\substack{+0.5 \\ -0.5}$    &  1.4 & $\substack{+0.3 \\ -0.2}$    &  2.7 & $\substack{+0.3 \\ -0.4}$    & 5.3 \\[2pt]
Per-emb 26 & 9(1)    &  3.4 & $\substack{+0.9 \\ -1.0}$    &  2.2 & $\substack{+0.9 \\ -0.7}$    &  3.6 & $\substack{+0.2 \\ -0.2}$    & 2.0 \\[2pt]
Per-emb 27 & 19.0(4) &    5 & $\substack{+1 \\ -1}$        &  1.4 & $\substack{+0.2 \\ -0.2}$    &  3.6 & $\substack{+0.3 \\ -0.2}$    & 3.0 \\[2pt]
Per-emb 33 & 8.3(8)  &    5 & $\substack{+3 \\ -1}$        &  1.5 & $\substack{+0.4 \\ -0.3}$    &  3.6 & $\substack{+0.5 \\ -0.5}$    & 4.9 \\[2pt]
Per-emb 44 & 32(7)   &  6.4 & $\substack{+0.2 \\ -0.2}$    & 1.27 & $\substack{+0.05 \\ -0.05}$  &  6.5 & $\substack{+0.5 \\ -0.4}$    & 6.7 \\[2pt]
Per-emb 63 & 1.9(4)  &    3 & $\substack{+1 \\ -2}$        &  1.6 & $\substack{+9.8 \\ -0.5}$    & \multicolumn{2}{c}{...} & 3.4 \\[2pt]
\hline
\end{tabular}
\tablefoot{Same structure as in Table~\ref{tbl:results}, except that the two columns to the right of the vertical line list the measured extents reported in the main paper, and the original extents reported by \citet{Jorgensen:2015kz}.}
\end{table*}

\section{Radiative transfer models of embedded protostars}
\label{sec:appB}

It is important to have a reliable estimate of the expected size of the \co emitting region to compare to the \co extents measured from real observations. The best way to address this is to use radiative transfer models to calculate synthetic observations, from which the sizes of the \co emitting regions can be measured. For one-dimensional power law envelope models, where the CO is assumed to be entirely in the gas-phase at temperatures >\SI{30}{K} and depleted by two orders of magnitude at temperatures <\SI{30}{K}, this yields an expected size of the CO emitting region of (see \citealt{Jorgensen:2015kz,Frimann:2016ke})
\begin{equation}
  \mathrm{FWHM}_\mathrm{avg} = \SI{0.77}{arcsec} \left(\frac{d}{\SI{235}{pc}}\right)^{-1} \left(\frac{L_\mathrm{bol}}{1\,L_\sun}\right)^{0.52},
  \label{eq:extent_orig}
\end{equation}
where $d$ is the distance to the source. The estimate is found to be relatively independent of the envelope mass, as the CO ice-line typically resides in the optically thin part of the envelope. A one-dimensional envelope model does not take into account any possible influence of the disk and outflow on the emission morphology. Furthermore, the analysis leading up to Eq.~\eqref{eq:extent_orig} also did not take into account the limited baseline coverage offered by a real interferometer or the noise inherent to such observations. Finally, the inclusion of disks and outflows will also influence the measured bolometric luminosities, which again will influence the expected size of the \co emitting region as a function of luminosity. In this Appendix, we use two-dimensional radiative transfer models (presented in Sect.~\ref{sec:Brad}) to study the influence of disks and outflows on the measured luminosities (Sect.~\ref{sec:Blum}) and on the sizes of the \co emitting regions (Sect.~\ref{sec:Bco}).

\subsection{Model description}
\label{sec:Brad}

The radiative transfer models are calculated using the dust radiative transfer code, \texttt{Hyperion} \citep{Robitaille:2011fc}, which uses the iterative monte carlo method of \citet{Lucy:1999wd} to calculate the dust temperatures of the models. The models are set up on a regular two-dimensional spherical grid, going from \SI{0.5}{AU} to \SI{8000}{AU} in the radial direction in 800 logarithmically spaced steps, and from $0$ to $\pi$ in the polar direction in 500 equally spaced steps. We assume the canonical gas-to-dust mass ratio of $100$ everywhere and use dust opacities from \citep{Ossenkopf:1994tq} corresponding to coagulated dust grains with thin ice-mantles at a density of $n_\mathrm{H2} \sim \SI{e6}{cm^{-3}}$ (OH5 dust). The temperatures in each model are calculated for ten different protostellar luminosities spaced equally in log-space between 1\,$L_\sun$ and 100\,$L_\sun$. The spectrum of the central source is assumed to be a black body with a temperature of \SI{4000}{K}.

\begin{table*}
\caption{Model parameters}
\label{tbl:parameters}
\centering
\begin{tabular}{l l l}
\hline\hline
Parameter & Value & Description \\
\hline
\multicolumn{3}{c}{Central Source} \\
\hline
$L_\mathrm{input}$             & 1.0, 1.7, 2.8, 4.6, 7.7, 12.9, 21.5, 35.9, 59.9, 100.0\,$L_\sun$ & Protostellar luminosity \\
$T_\mathrm{eff}$               & \SI{4000}{K} & Effective temperature \\
\hline
\multicolumn{3}{c}{Disk} \\
\hline
$M_\mathrm{disk}$              & 0.0, 0.01, 0.1, 0.3\,$M_\sun$ & Disk mass \\
$\gamma$                       & $1$ & Disk power law index\\
$r_\mathrm{min}^\mathrm{disk}$ & \SI{0.5}{AU} & Disk inner radius \\
$r_\mathrm{max}^\mathrm{disk}$ & \SI{100}{AU} & Disk outer radius \\
$r_0^\mathrm{disk}$            & \SI{100}{AU} & Disk reference radius \\
$h_0$                          & \SI{10}{AU}  & Disk scale hight at $r_0$ \\
$\beta$                        & $1.3$ & Disk flaring index \\
\hline
\multicolumn{3}{c}{Envelope} \\
\hline
$M_\mathrm{env}$               & 0.3, 1.0, 3.0\,$M_\sun$ & Envelope mass \\
$p$                            & 1.5, 2.0 & Envelope power law index \\
$r_\mathrm{min}^\mathrm{env}$  & \SI{0.5}{AU} (\SI{5}{AU} for envelope-only models) & Envelope inner radius \\
$r_\mathrm{max}^\mathrm{env}$  & \SI{8000}{AU} & Envelope outer radius \\
$r_0^\mathrm{env}$             & \SI{0.5}{AU} & Envelope reference radius \\
\hline
\multicolumn{3}{c}{Bipolar Cavity} \\
\hline
$d$                            & $1.5$ & Cavity wall power law index \\
$\rho_0^\mathrm{cavity}$       & \SI{1.67e-20}{g.cm^{-3}} & Cavity density \\
$r_0^\mathrm{cavity}$          & \SI{1000}{AU} & Cavity reference radius \\
$\theta_0^\mathrm{cavity}$     & \ang{0}, \ang{10}, \ang{20} & Half opening angle \\
\hline
\multicolumn{3}{c}{Miscellaneous} \\
\hline
$\cos i$                       & 0.0, 0.1, 0.2, 0.3, 0.4, 0.5, 0.6, 0.7, 0.8, 0.9, 1.0 & Viewing angles \\
$\mathcal{R}_\mathrm{gd}$      & 100 & Gas-to-dust mass ratio \\
$n_\mathrm{c18o}^\mathrm{gas}/n_\mathrm{H}$   & \num{2e-7} if $T_\mathrm{dust}$ > $T_\mathrm{sub}$; \num{2e-9} otherwise & C$^{18}$O gas-phase abundance \\
\hline
\end{tabular}
\end{table*}

The physical structure of each model is made up of an axissymmetric disk, an envelope, and an outflow or some combination thereof. The physical structure of each model component is described in the following, with model parameters listed in Table~\ref{tbl:parameters}. The disk density is given by
\begin{equation}
  \rho_\mathrm{disk} \left(r,\theta \right) = \rho_0^\mathrm{disk} \left(\frac{r \sin \theta}{r_0^\mathrm{disk}}\right)^{-\gamma-\beta} \exp \left[ -\frac{1}{2} \left( \frac{r \cos \theta}{h\left(r,\theta\right)}\right)^2 \right], \nonumber
\end{equation}
where $r$ is the radial coordinate and $\theta$ is the polar angle measured from the rotation axis. The pressure scale height, $h\left(r,\theta\right)$, is given by
\begin{equation}
  h\left(r,\theta\right) = h_0 \left(\frac{r \sin \theta}{r_0^\mathrm{disk}}\right)^\beta. \nonumber
\end{equation}
The scale parameter, $\rho_0^\mathrm{disk}$, is not set directly but is calculated by integrating the disk model over the three spatial dimensions, setting the result equal to the disk mass, $M_\mathrm{disk}$. The disk power law index, $\gamma$, and the flaring index, $\beta$, are fixed at $1$ and $1.3$, appropriate for young disks \citep{Williams:2011js}. The disk's inner and outer radius, $r_\mathrm{min}^\mathrm{disk}$ and $r_\mathrm{max}^\mathrm{disk}$, are fixed at \SI{0.5}{AU} and \SI{100}{AU} respectively, while the disk mass can take three different values, 0.01, 0.1, 0.3\,$M_\sun$, which are chosen to be representative of the range of compact masses listed in Table~\ref{tbl:continuum_results}. A set of models with no disk (disk mass of 0.0\,$M_\sun$) are also calculated.

The envelope density is given by
\begin{equation}
  \rho_\mathrm{env} \left(r \right) = \rho_0^\mathrm{env} \left(\frac{r}{r_0^\mathrm{env}}\right)^{-p}, \nonumber
\end{equation}
where again the scale parameter, $\rho_0^\mathrm{env}$, is set by integrating over the three spatial dimensions and equating the result to the envelope mass, $M_\mathrm{env}$. The envelope power law index, $p$, can take on two values, $1.5$ and $2.0$, appropriate for an isothermal singular sphere in either free-fall or hydrostatic equilibrium \citep{Shu:1977ef}. The envelope's inner and outer radius, $r_\mathrm{min}^\mathrm{env}$ and $r_\mathrm{max}^\mathrm{env}$, are fixed at \SI{0.5}{AU} and \SI{8000}{AU} respectively. $M_\mathrm{env}$ can take on values of 0.3, 1.0, and 3.0\,$M_\sun$ chosen to be representative of the range of envelope masses listed in Table~\ref{tbl:continuum_results}.

The walls of the outflow cavity are described by
\begin{equation}
  r \cos \theta = r_0^\mathrm{cavity} \cos \theta_0^\mathrm{cavity} \left(\frac{r \sin \theta}{r_0^\mathrm{cavity} \sin \theta_0^\mathrm{cavity}} \right)^d. \nonumber
\end{equation}
Inside the cavity walls the density is held constant at $\rho_0^\mathrm{cavity}$, except that the density is capped so that it will never exceed the envelope density. $d$ is fixed to $1.5$ in our models, while the cavity opening \emph{half}-angle, $\theta_0^\mathrm{cavity}$, takes on values of \ang{10} and \ang{20} (measured at a reference distance, $r_0^\mathrm{cavity}$, of \SI{1000}{AU}) appropriate for Class~0 and~I protostars (e.g.\ \citealt{Arce:2006dt}). A set of models without any outflow ($\theta_0^\mathrm{cavity} = \ang{0}$) is also calculated. The density, $\rho_0^\mathrm{cavity}$, is fixed at \SI{1.67e-20}{g.cm^{-3}}, corresponding to a Hydrogen number density of \SI{e4}{cm^{-3}}.

A total of 630 radiative transfer models are calculated of which 360 consist of both disk, envelope, and outflow components (henceforth denoted full models); 120 consist of envelope and outflow components; 90 of disk and envelope components; and 60 only of an envelope component. Instead of the standard two-dimensional grid, the envelope-only models are calculated on a one-dimensional radial grid extending from \SIrange{5}{8000}{AU} in \num{800} logarithmically spaced intervals. The models span the entire parameter space listed in Table~\ref{tbl:parameters}, except for the disk+envelope models, which have only been calculated for an envelope power law index, $p$, of \num{1.5} to save computing time.

Synthetic spectral energy distributions (SEDs) and C$^{18}$O\,\mbox{2--1} moment zero maps are calculated for each of the radiative transfer models, using the radiative transfer code \texttt{RADMC-3D}\footnote{\url{http://www.ita.uni-heidelberg.de/~dullemond/software/radmc-3d/}} (see \citealt{Dullemond:2004iy} for a description of the two-dimensional version of this code). Except for the spherically symmetric envelope-only models, the synthetic observables are calculated at 11 different viewing angles, $i$, distributed uniformly in $\cos i$ between $i$\,=\,\ang{0} (pole-on) and $i$\,=\,\ang{90} (edge-on). Synthetic observables of the envelope-only models are only calculated at one viewing angle.

\begin{figure*}
\includegraphics[width=18cm]{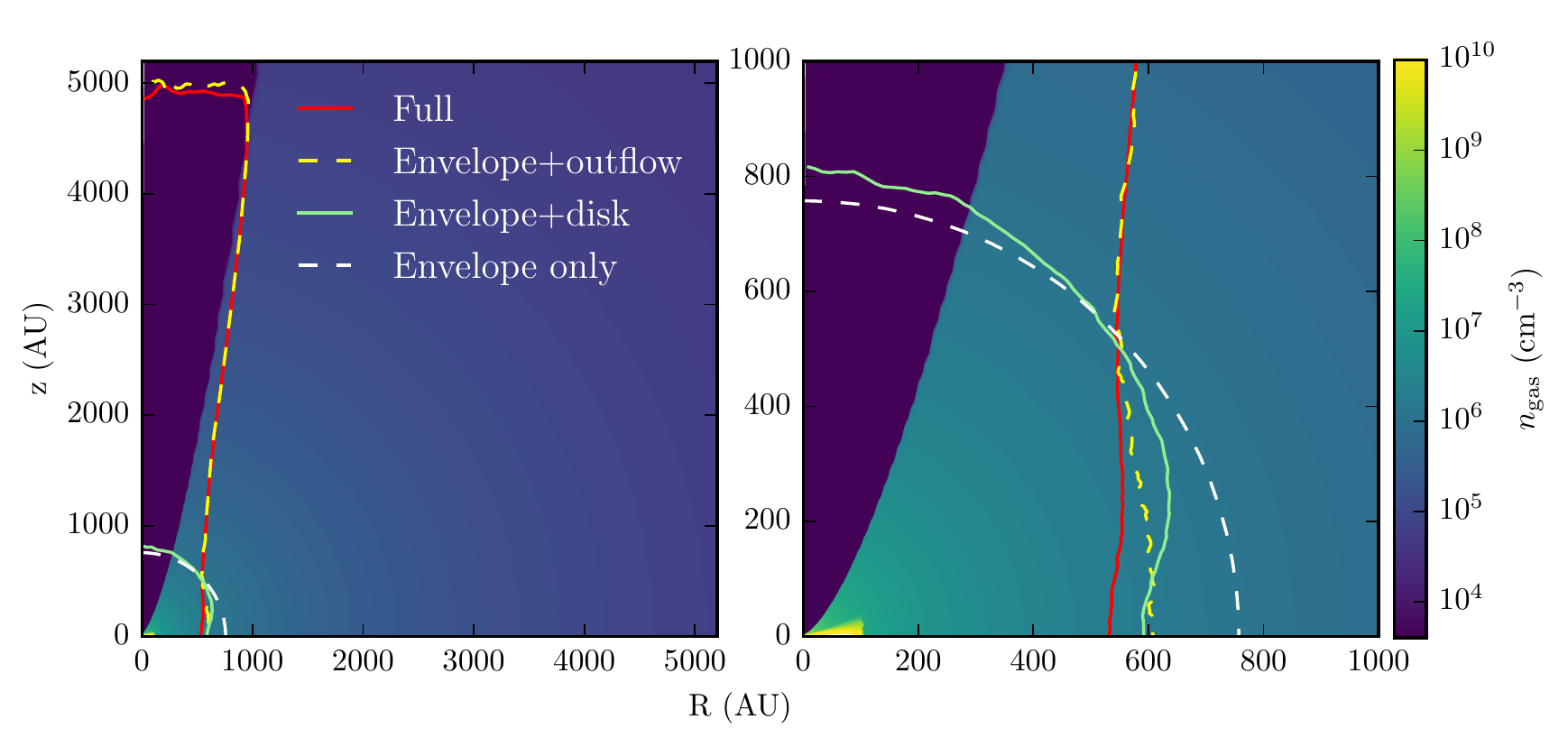}
   \caption{Density cross section of one of the full radiative transfer models. Both panels show the same model, but the right panel is zoomed in relative to the left. The model has parameters $M_\mathrm{disk}$\,=\,$0.1\,M_\sun$, $M_\mathrm{env}$\,=\,$1\,M_\sun$, $p$\,=\,1.5, $\theta_0^\mathrm{cavity}$\,=\, \ang{20}, and $L_\mathrm{input}$\,=\,$12.9\,L_\sun$. The figure also shows the positions of the CO ice lines at \SI{28}{K} for the full model, as well as for the equivalent models that do not include the disk and/or outflow.}
   \label{fig:contour}
\end{figure*}

The synthetic SEDs are calculated at wavelengths extending from \SIrange{0.1}{1000}{\micro\metre} in \num{100} logarithmically spaced intervals. The aperture used for calculating the SED is centred on the protostar and has a diameter of \SI{3525}{AU}, corresponding to \ang{;;15} at a distance of \SI{235}{pc}. To calculate the \co moment zero maps, we assume local thermodynamic equilibrium and that gas and dust temperatures are equal. Furthermore, we assume a canonical CO abundance of \num{e-4}\,$n_\mathrm{co}/n_\mathrm{H}$ and a $^{16}$O/$^{18}$O ratio of 500. At temperatures $>T_\mathrm{sub}$, CO is assumed to be entirely in the gas phase, while at temperatures $<T_\mathrm{sub}$, the abundance is depleted by two orders of magnitude to emulate freeze-out \citep{Jorgensen:2005df}. We adopt two sublimation temperatures: $T_\mathrm{sub} = \SI{28}{K}$, which is applied to all models, and $T_\mathrm{sub} = \SI{21}{K}$, which is applied to one set of full models.

Figure~\ref{fig:contour} shows the density cross section of one of the full models, along with the location of the CO ice line at $T_\mathrm{dust}$\,=\,\SI{28}{K}. For comparison, the figure also shows the locations of the ice lines for the equivalent models that do not include the disk and/or outflow. We remark that the purpose at hand is not to do a precise modelling of disks, envelopes, and outflows, but to investigate the influence of disks and outflows on the measured bolometric luminosities and on the morphologies of the C$^{18}$O\,2--1 emission.

\begin{figure*}
\sidecaption
  \includegraphics[width=12cm]{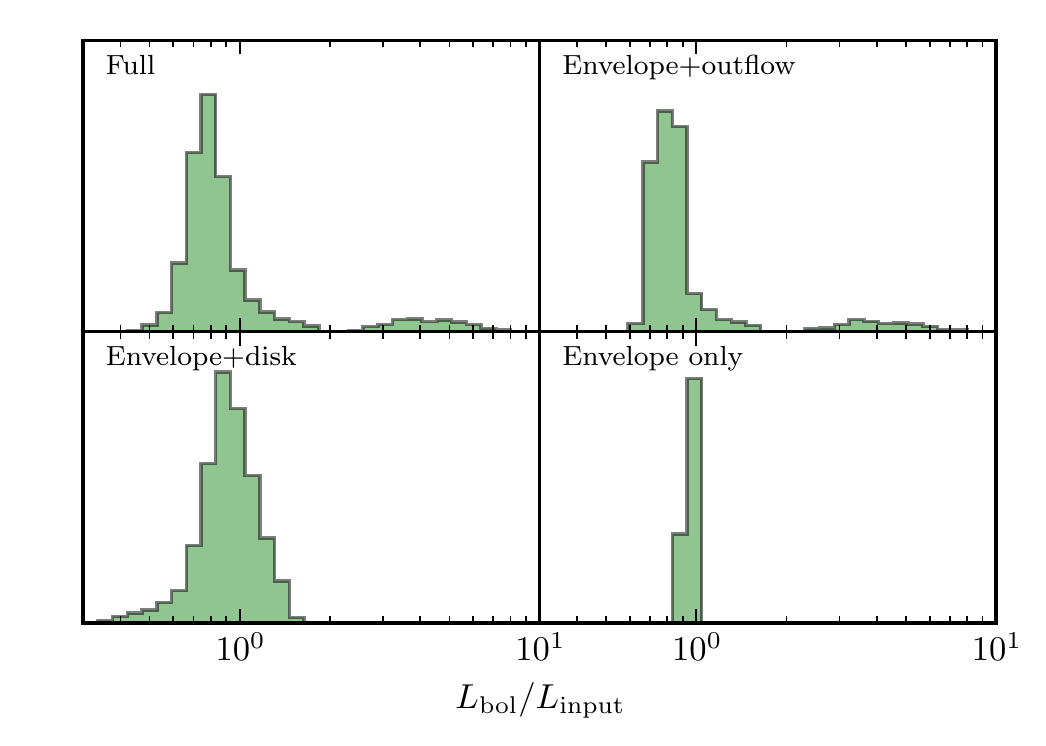}
     \caption{Distribution of $L_\mathrm{bol}/L_\mathrm{input}$ for the different sets of models. Basic statistics of the distributions are listed in Table~\ref{tbl:lumdist}.}
     \label{fig:lumdist}
\end{figure*}

\begin{table}
\caption{$L_\mathrm{bol}/L_\mathrm{input}$ distribution statistics.}
\label{tbl:lumdist}
\centering
\begin{tabular}{l S[table-format=1.2] S[table-format=1.2] S[table-format=1.2] S[table-format=1.2] S[table-format=1.2]}
\hline\hline
 & {Min} & {Max} & {Mean} & {Median} & {SD} \\
\hline
Full             & 0.45 & 7.41 & 0.83 & 0.80 & 0.19 \\
Envelope+outflow & 0.64 & 7.43 & 0.84 & 0.82 & 0.15 \\
Envelope+disk    & 0.35 & 1.52 & 0.93 & 0.92 & 0.20 \\
Envelope only    & 0.85 & 1.00 & 0.95 & 0.96 & 0.03 \\
\hline
\end{tabular}
\tablefoot{Min, max, mean, median, and standard deviation of the distributions plotted in Fig.~\ref{fig:lumdist}. The mean, median, and standard deviation exclude $L_\mathrm{bol}/L_\mathrm{input}$\,>\,2 to eliminate viewing angles along the outflow.}
\end{table}

\subsection{Luminosities}
\label{sec:Blum}

Photons escape more easily through low-density regions, where they are subjected to fewer scattering and absorption events, than though high-density regions, where the number of dust gains they can interact with is large. Consequently, one can expect the measured luminosity towards a source containing a disk or an outflow to depend on the viewing angle, whereas the measured luminosity towards a spherically symmetric source will be independent of the viewing angle.

Figure~\ref{fig:lumdist} shows the distribution of $L_\mathrm{bol}/L_\mathrm{input}$ for the different sets of radiative transfer models, where $L_\mathrm{bol}$ has been calculated by integrating over the synthetic SEDs. The most striking features are the narrowness of the envelope-only distribution, which is expected since the models are spherically symmetric, and the tail of high $L_\mathrm{bol}/L_\mathrm{input}$ in the outflow models, which corresponds to viewing angles where one looks into the outflow.

\begin{figure*}
\sidecaption
  \includegraphics[width=12cm]{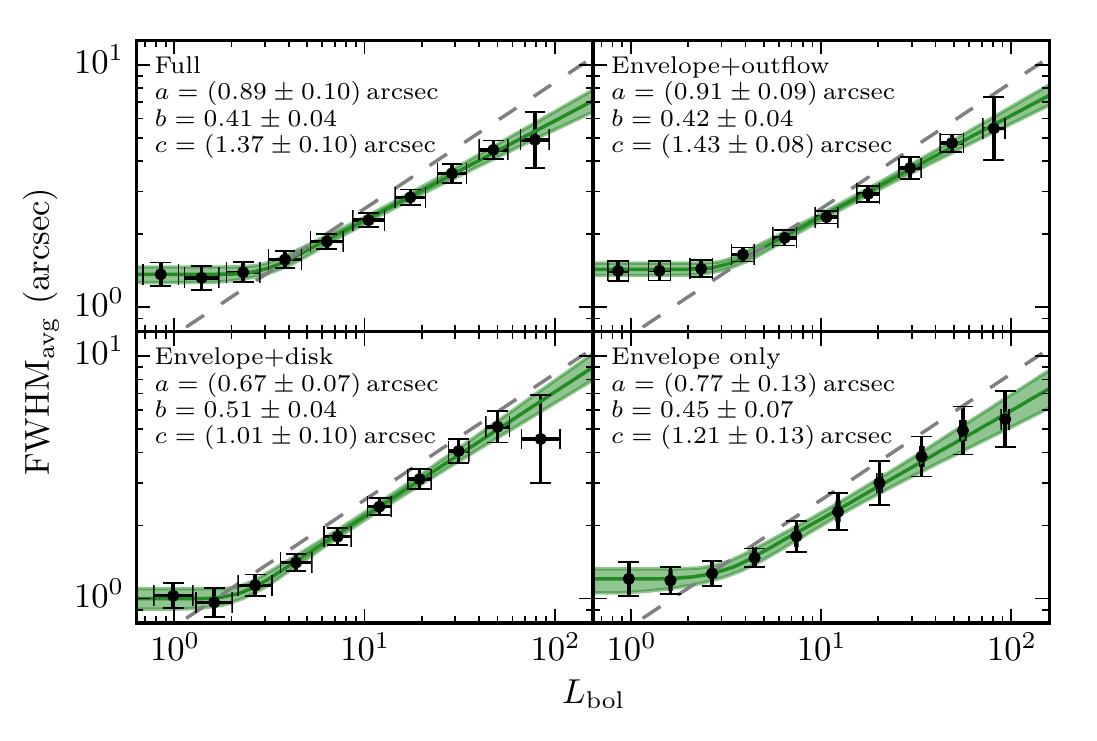}
     \caption{Size of the \co emitting region as function of the measured luminosity, $L_\mathrm{bol}$, for the different sets of models. The errorbars indicate binned results, and the best fitting model, along with the 1$\upsigma$ uncertainty region is shown in green. Equation~\eqref{eq:extent_orig} is shown as the grey dashed line.}
     \label{fig:analytical_extent}
\end{figure*}

Table~\ref{tbl:lumdist} presents basic statistics of the four distributions in Fig.~\ref{fig:lumdist}. As evident from the numbers, the most extreme deviations from the input luminosity occur in models with outflows, when the viewing angle is aligned with the outflow. This extreme situation is comparatively uncommon, and we therefore exclude viewing angles along the outflow when calculating the mean, median and standard deviation of the distributions. On average, the measured luminosities are smaller than the input luminosities because edge-on viewing angles are more likely than face-on viewing angles and because of the finite size of the aperture, which excludes extended emission. The magnitude of the latter effect can be estimated by looking at the envelope-only models, which deviate at most by \SI{15}{\percent} from the input luminosity and typically by much less. The average luminosity is reduced by a larger factor  when including an outflow than it is when including a disk. This stems from the fact that a large fraction of the radiation escapes through the outflow, reducing the bolometric luminosity measured through the remaining viewing angles. On the other hand, models that contain disks show broader distributions than models without (again, excluding viewing angles that are along the outflow). This stems from the fact that the measured bolometric luminosity varies more strongly with viewing angle for models that contain a disk relative to models that do not.

While extreme situation do occur, we find that the measured bolometric luminosities are, on average, reduced by \SI{20}{\percent} relative to the input luminosity, and that deviations from the input luminosity larger than a factor of 2 are rare. The notable exception regards systems that are viewed through the outflow, where the measured bolometric luminosity may be enhanced by as much as a factor of 7 relative to the input luminosity. We do not believe this to be a great concern for the objects in our observed sample, since this case is expected to be rare, and because the majority of the observed systems have identified outflows, which suggests edge-on geometries. Our results are also in qualitative agreement with previous studies (e.g.\ \citealt{Whitney:2003ke,Offner:2012bc}).

\subsection{C$^{18}$O emitting region}
\label{sec:Bco}

We wish to calculate the sizes of the \co emitting regions in the radiative transfer models and evaluate how they depend on the protostellar luminosity by fitting them to a functional relationship like that in Eq.~\eqref{eq:extent_orig}. As a first step, the synthetic moment zero maps are Fourier transformed and sampled in the $(u,v)$-plane using the $(u,v)$-coverage of Per-emb~53 as a template. Noise with a standard deviation of \SI{4}{Jy.km.s^{-1}} is subsequently added to the synthetic $(u,v)$-data, again based on estimates of the noise in the Per-emb~53 data. Two-dimensional Gaussians are fitted to the synthetic $(u,v)$-data. The MCMC method presented in Sect.~\ref{sec:fitting} is not used to fit the data, as it would be impractical given the large number of models; instead we use a standard non-linear fitting algorithm and ensure that the fit converges to a reasonable result by giving good starting guesses. Finally, the measured sizes of the C$^{18}$O emitting regions are binned according to their input luminosity and the set of models they belong to. The binned data are fitted to a functional relationship of the form
\begin{equation}
  \mathrm{FWHM}_\mathrm{avg}\left(x\right) = \begin{cases} \mathrm{FWHM}\left(x\right) & \textit{if} \quad \mathrm{FWHM}\left(x\right) > c \\
  c & \textit{if} \quad \mathrm{FWHM}\left(x\right) \leq c\end{cases}, \nonumber
\end{equation}
where
\begin{equation}
  \mathrm{FWHM}\left(x\right) = a \left(\frac{x}{1\,L_\sun}\right)^{b}. \nonumber
\end{equation}
The constant, $c$, is included to account for the fact that noise and $(u,v)$-coverage limit the sizes of the extents that can be measured, and $c$ is therefore an estimate of that limit. Note that because the fitting is done directly in the $(u,v)$-plane the value of $c$ can be smaller than the size of the synthesised beam.

The results of the fitting procedure for the $T_\mathrm{sub} = \SI{28}{K}$ models are presented in Fig.~\ref{fig:analytical_extent}. Although the fitted parameters vary somewhat between the different sets of models, they are generally consistent with each other within 2$\upsigma$. This suggests that the inclusion of disks and outflows has little influence on the location of the CO ice line, something that is also confirmed in Fig.~\ref{fig:contour}. Naturally, the CO ice line extends to large radii inside the outflows, but since the densities are so low, this has little effect on the emission. Comparing to Eq.~\eqref{eq:extent_orig}, we see that the power law exponents, $b$, derived here are, on average, somewhat flatter. We attribute this difference to the fact that the synthetic observables used have use a realistic $(u,v)$-coverage, and have had noise added to them, making them resemble real observations more. The expression that is used to predict the expected size of the \co emitting region in the main paper, is the one for the full models, and is given as
\begin{equation}
  \mathrm{FWHM}_\mathrm{avg} = \SI{0.89}{arcsec} \left(\frac{d}{\SI{235}{pc}}\right)^{-1} \left(\frac{L_\mathrm{bol}}{1\,L_\sun}\right)^{0.41}. \nonumber
\end{equation}
We also fit the full models calculated for a sublimation temperature of \SI{21}{K} to the functional relationship. The result is equal to that for a sublimation temperature of \SI{28}{K}, except that the angular coefficient, $a$, is increased to \SI[separate-uncertainty]{1.64(10)}{arcsec}.

\bibliographystyle{aa} 
\bibliography{citations} 

\end{document}